\documentclass[lettersize,journal]{IEEEtran}
\usepackage{amsmath,amsfonts}
\usepackage{array}
\usepackage[caption=false,font=normalsize,labelfont=sf,textfont=sf]{subfig}
\usepackage{textcomp}
\usepackage{stfloats}
\usepackage{url}
\usepackage{verbatim}
\usepackage{graphicx}
\usepackage{cite}
\usepackage{enumitem}
\usepackage[ruled,vlined]{algorithm2e}
\usepackage{tabularx}
\usepackage{booktabs}
\usepackage{multirow}
\usepackage[colorlinks=true, linkcolor=blue, citecolor=blue, urlcolor=blue]{hyperref}
\usepackage{orcidlink}
\begin{document}

\title{HMS-SCP: Task-Oriented Multi-Scale Semantic Communication for V2X Cooperative Perception}

\author{Chun-Yeow Yeoh,~\IEEEmembership{Senior~Member,~IEEE,\orcidlink{0000-0003-0555-8049}}
        Chee Keong Tan,~\IEEEmembership{Senior-Member,~IEEE,\orcidlink{0000-0001-5551-1017}}
        Joanne~Mun-Yee~Lim,~\IEEEmembership{Senior-Member,~IEEE,\orcidlink{0000-0002-1326-8634}}
        and~Heng-Siong Lim,~\IEEEmembership{Senior~Member,~IEEE,\orcidlink{0000-0002-8346-6345}}
        % <-this % stops a space
        
\thanks{This work has been submitted to the IEEE for possible publication. Copyright may be transferred without notice, after which this version may no longer be accessible.}        
\thanks{Chun-Yeow Yeoh and Chee Keong Tan are with the School of Information Technology, Monash University Malaysia, Malaysia (e-mail: chunyeow.yeoh@monash.edu; tan.cheekeong@monash.edu).}
\thanks{Joanne Mun-Yee Lim is with the Department of Electrical and Robotics Engineering, School of Engineering, Monash University Malaysia, Malaysia (e-mail: Joanne.Lim@monash.edu).}
\thanks{Heng Siong Lim is with the Faculty of Engineering and Technology, Multimedia University, Malaysia (e-mail: hslim@mmu.edu.my).}
}

\markboth{Journal of \LaTeX\ Class Files,~Vol.~XX, No.~X, XXX~2026}%
{Shell \MakeLowercase{\textit{et al.}}: Task-Oriented Multi-Scale Semantic Communication for V2X Cooperative Perception}


\maketitle

\begin{abstract}
Cooperative perception enables vehicles and infrastructure to exchange sensor data via Vehicle-to-Everything (V2X) communication, extending sensing coverage beyond occlusions and mitigating blind spots. While critical for autonomous driving and safety, practical deployments often rely on bandwidth-efficient late fusion. Recently, intermediate fusion has emerged as a promising approach for an optimal bandwidth-accuracy trade-off. However, in dense urban environments, cumulative bandwidth demands can overwhelm network capacity, potentially compromising safety-critical Cooperative Intelligent Transport Systems (C-ITS) functions. To alleviate these problems, this paper proposes Hierarchical Multi-Scale Semantic-Aware Cooperative Perception (HMS-SCP), a robust noise-resilient and bandwidth-efficient framework for task-oriented semantic communication in cooperative perception. HMS-SCP employs a spatial importance predictor to identify task-relevant grid elements at each scale, which are then directly mapped into complex-valued symbols for Joint Source-Channel Coding (JSCC). Unlike prior methods that rely on high-dimensional symbol projections for robustness, HMS-SCP exploits structural semantic redundancy across multiple scales to enhance resilience against channel noise, while maintaining an ultra-low symbol rate. This design significantly reduces bandwidth consumption and mitigates network congestion in high-density vehicular environments. Extensive evaluations on the simulated OPV2V and real-world DAIR-V2X datasets demonstrate that HMS-SCP effectively prevents performance collapse under severe Rayleigh fading and extreme compression ratio, maintaining high-confidence far-field detection with a real-time latency of below 16~ms, well within the safety-critical thresholds for dynamic V2X environments.
\end{abstract}

    
\section{Introduction}
\IEEEPARstart{R}{oad} traffic accidents claim approximately 1.19 million lives globally each year, making them the leading cause of death among children and young adults aged 5 to 29 years\cite{WHO2023RoadSafety}, according to World Health Organization (WHO). Addressing this urgent global transportation safety challenge has long motivated the International Telecommunication Union (ITU) to promote Intelligent Transport Systems (ITS)\cite{9779322}. ITS have evolved from basic traffic monitoring using induction loops to situational roadway analysis and interactive driver assistance. Beyond these driver assistance functions, the next generation of ITS places greater emphasis on cooperative perception, prediction, decision making, and control through V2X communication to enable autonomous driving. 
\begin{figure}[t]
    \centering
    \includegraphics[width=0.485\textwidth]{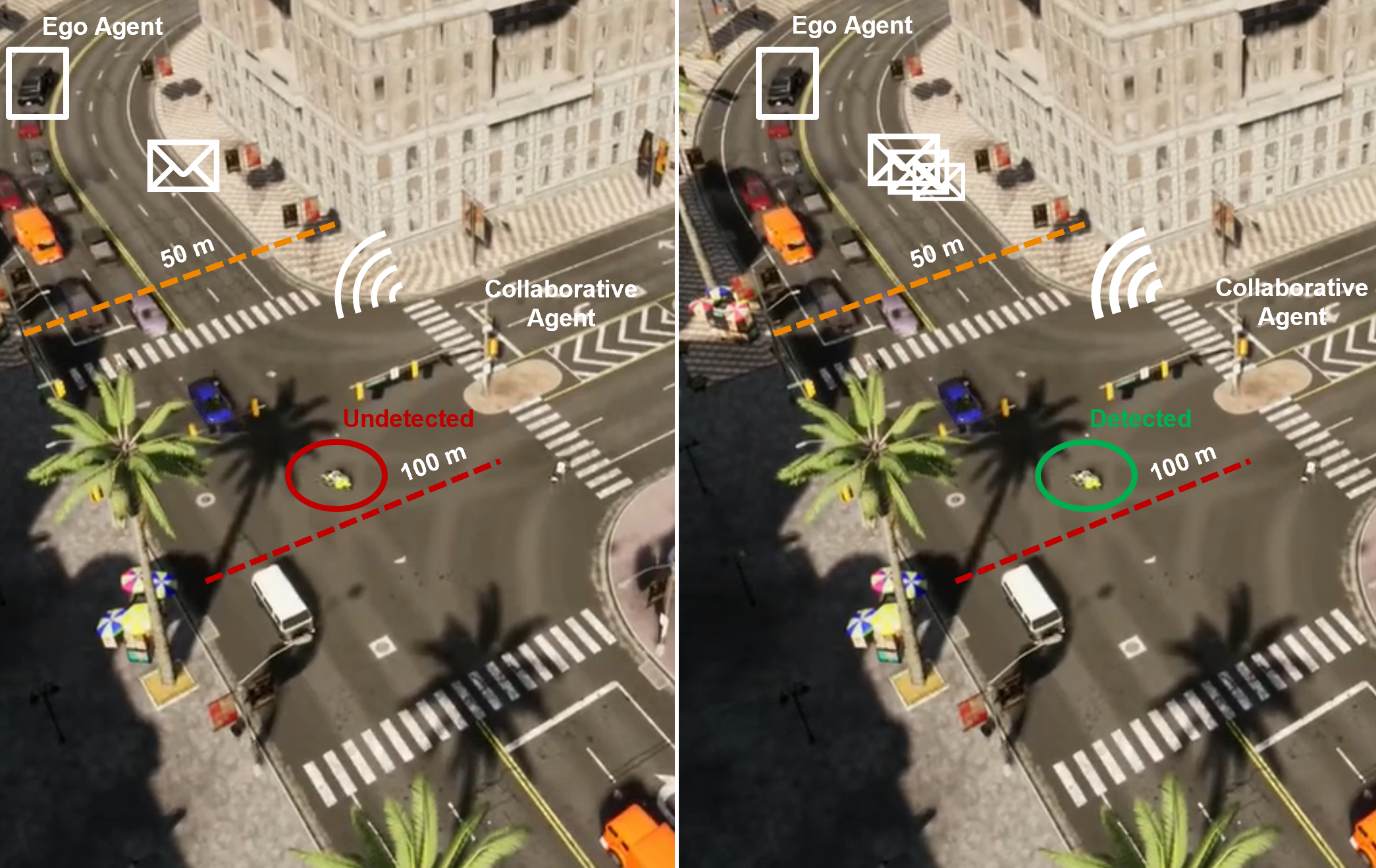} 
    \caption{Cooperative perception performance at extreme compression ratio ($\eta = 0.01$) and severe Rayleigh channel noise ($SNR = 0$ dB). The left panel illustrates the single-scale baseline which fails to resolve distant hazards beyond the 50~m mark. The right panel demonstrates our proposed hierarchical HMS-SCP framework, which successfully detects the 100~m hazard by preserving hierarchical features. This structural resilience effectively extends the ITS safety horizon, ensuring reliable perception under V2X communication constraints.}
    \label{fig:teaser_image}
\end{figure}
Conventional vehicular perception remains fundamentally constrained by their dependence on ego-centric sensor suites, including cameras, millimeter-wave (mmWave) radar, and Light Detection and Ranging (LiDAR). Although LiDAR has become a cornerstone technology for high-precision three-dimensional (3D) localization and environmental perception, single-vehicle sensing systems remain inherently vulnerable to critical performance degradation in complex driving environments. These limitations arise primarily from line-of-sight (LoS) occlusions, diminishing point-cloud density at extended sensing ranges, and the inability to perceive objects beyond physical obstructions. To address these challenges, recent advances in V2X communication, aligned with the C-ITS paradigm\cite{lu2018c}, have enabled the integration of heterogeneous sensor observations from distributed agents and roadside infrastructure to enhance sensing robustness, mitigates occlusion-induced failures, and extends the effective perception horizon beyond the physical constraints of individual vehicles\cite{10375912, PalladinAndBruckerLRS4Fusion}. 

State-of-the-art (SOTA) V2X technologies primarily include ETSI ITS-G5 and Cellular V2X (C-V2X). ETSI ITS-G5, based on IEEE 802.11p, enables low-latency ad hoc vehicular communication in the 5.9 GHz ITS spectrum. In practice, it typically operates over 10 MHz channels to achieve nominal physical-layer data rates up to 6~Mbps under favorable link conditions\cite{10056390}. In parallel, C-V2X that standardized under 3GPP Rel. 14/16, extends vehicular connectivity through direct sidelink (PC5) and network-assisted modes\cite{8581518}. Supporting flexible 10–20~MHz bandwidths, it achieves throughput of 6–12~Mbps depending on channel conditions, modulation, and resource allocation. These communication constraints impose a critical bottleneck for cooperative perception, where multiple agents must exchange perception information in real time under highly dynamic traffic conditions. Consequently, recent studies\cite{11144483} have focused on reducing transmission overhead through spatial sparsification, feature compression, and channel-efficient intermediate fusion strategies\cite{10588382} to better align cooperative perception workloads with practical V2X bandwidth limitations. However, most existing frameworks assume idealized or quasi-static communication links, overlooking perception degradation from physical-layer noise, multipath fading, and packet corruption. This disconnect between algorithmic design and realistic wireless channel behavior remains a primary barrier to real-world deployment\cite{11092441, Where2comm:22, xu2022v2xvit}.

To bridge this gap, recent advances in semantic communication have progressively transformed cooperative perception from raw data exchange to task-oriented semantic transmission\cite{weaver1949recent}, offering a more communication-efficient and channel-resilient solution for vehicular networks affected by severe and time-varying wireless impairments\cite{10405254, 11355867, 11153454}. Rather than conveying dense feature tensors, existing approaches prioritize importance-aware feature selection and the exchange of compact intermediate representations, such as learned latent features, object-level descriptors, or scene semantics, that are most relevant to downstream perception and cooperative fusion tasks. In this context, a deep Joint Source-Channel Coding (JSCC)-based architecture has emerged as a promising framework to improve robustness under bandwidth constraints and low signal-to-noise ratio (SNR) conditions. By jointly optimizing source compression and channel transmission in an end-to-end manner, JSCC-based systems exhibit graceful performance degradation as channel quality deteriorates, in sharp contrast to the abrupt cliff effect commonly observed in conventional separate source channel coding\cite{6773024, 11192484, 10679082}. 

However, based on our observation, the existing semantic encoding process is generally restricted to a single-scale feature map, as exemplified by recent studies\cite{10405254, 11355867, 10810363}. As shown in Fig.~\ref{fig:teaser_image}, this design introduces a fundamental limitation to the perception performance, particularly for distant actors that appear at smaller scales in the feature hierarchy\cite{10965738, 10027465, lu2024an}. This results in a shorter safety horizon, which is particularly dangerous at high speeds. In an ITS context, safe navigation at 100~km/h requires a perception horizon beyond 250~m\cite{PalladinAndBruckerLRS4Fusion} to maintain safe reaction times of approximately 6 to 9 seconds. Ego-centric perception becomes fundamentally insufficient at higher speeds, suffering a precipitous performance collapse beyond a 50~m threshold. To bridge this critical 50-250~m safety gap within stringent 6–12~Mbps V2X bandwidth constraints, we propose the collaborative HMS-SCP framework. By extracting and jointly optimizing hierarchical semantic representations, our approach replaces traditional channel coding redundancy\cite{11355867} with multi-scale semantic redundancy. This paradigm shift improves transmission efficiency and noise resilience, achieving the optimal bandwidth-accuracy trade-off necessary for high-speed autonomous navigation. 

The main contributions of this work are as follows:
\begin{enumerate}[label={[\arabic*]}, topsep=0pt]
    \item We pioneer a hierarchical multi-scale semantic communication paradigm for cooperative perception, moving beyond the conventional single-scale transmission framework. By jointly exploiting three complementary feature levels through deep JSCC, the proposed architecture preserves both fine-grained spatial details and high-level contextual semantics, substantially extending the reliable perception horizon under severe occlusion and wireless impairments. 
    \item We introduce a fundamentally new structural-redundancy principle for channel robustness to replace the brute-force high-dimensional symbol expansion adopted in prior semantic communication methods. Instead of relying on excessive symbol repetition, our framework leverages cross-scale semantic complementarity to achieve strong resilience against Rayleigh fading and low-SNR conditions while maintaining ultra-low transmission overhead.
    \item We develop an importance-aware rate-controlled semantic transmission mechanism that dynamically selects the most task-critical spatial features across multiple scales and maps each selected location into only a single complex-valued symbol. This enables an ultra-efficient transmission ratio of $R=0.01$, delivering up to 256× higher symbol efficiency than SOTA baselines without sacrificing perception accuracy.
    \item We establish a practical modular training strategy for next-generation V2X perception systems, allowing the proposed semantic communication layers to be seamlessly integrated with pretrained cooperative perception backbones and detection heads without full retraining. This significantly reduces the complexity of the deployment while preserving pretrained perception capability.
    \item We provide the comprehensive evidence that hierarchical semantic communication can simultaneously achieve near-upper-bound perception accuracy, real-time latency, and extreme bandwidth efficiency under adverse channel conditions. Extensive experiments on the simulated OPV2V and real-world DAIR-V2X benchmarks demonstrate superior robustness under Additive White Gaussian Noise (AWGN)  and Rayleigh fading, with particularly large gains in the safety-critical range ($50\text{--}100\text{m}$) where existing methods collapse.
\end{enumerate}

\section{Related Works}
\subsection{V2X based Cooperative Perception}

While single-vehicle Bird's-Eye-View (BEV) perception provides a convenient unified representation for downstream detection, these ego-centric models\cite{10160968, 10791908} remain fundamentally constrained by LoS occlusions, sparse long-range sensing, and viewpoint limitations. To overcome these deficiencies, multi-agent cooperative perception frameworks leveraging Vehicle-to-Vehicle (V2V) and Vehicle-to-Infrastructure (V2I) communications have emerge as a robust alternative, enabling the aggregation of complementary spatial observations across distributed agents. By fusing sensor data from across-agents, these systems significantly mitigate the unreliability of individual sensors in challenging environments. From a system design perspective, cooperative perception frameworks can be broadly categorized into early, intermediate, and late fusion\cite{9228884, 11272131}. Early fusion provides a theoretical performance upper bound through raw data exchange but remains generally impractical due to extreme bandwidth demands. Conversely, late fusion minimizes network load by sharing only final outputs, such as bounding boxes, though it often lacks the semantic depth required to resolve complex occlusions. 

More recently, intermediate fusion for cooperative perception has emerged as a dominant paradigm due to its optimal trade-off between perception accuracy and communication efficiency\cite{10588382}. Hu et al. proposed Where2comm\cite{Where2comm:22}, an intermediate fusion framework that introduces a spatial confidence map to selectively transmit perceptually critical regions, significantly reducing communication bandwidth while maintaining high perception performance. Building on the principle of selective communication, Yang et al. proposed How2comm\cite{yang2023how2comm}, which employs a mutual-information-aware communication mechanism with spatial–channel feature filtering to sparsify transmitted features, enabling agents to transmit only informative perception features. In a different direction, Hu et al. proposed CodeFilling\cite{YueCodeFilling:CVPR2024}, which reduces communication bandwidth by encoding perception features using a shared vector-quantized codebook and transmitting compact code indices for informative tokens, thereby enabling highly compressed communication for cooperative perception. Subsequently, Tao et al.\cite{11127818} proposed enabling an ego vehicle to proactively prioritize specific directions of interest and intelligently reallocate limited bandwidth to those vital areas using a direction-aware selective attention mechanism to enhance local directional perception. 

Despite the success of intermediate fusion in reducing bandwidth\cite{11092441, Where2comm:22, xu2022v2xvit, yang2023how2comm, YueCodeFilling:CVPR2024, 10160546, zhou2025v2xpnp}, existing frameworks remain fundamentally limited in three key aspects. First, they largely assume ideal or quasi-static communication channels, making them vulnerable to performance degradation under realistic wireless impairments such as fading, noise, and packet corruption. Second, they treat feature transmission primarily as a spatial pruning or compression problem, which reduces bandwidth but still transmits raw feature bits that may contain high-frequency noise or task-irrelevant background. Third, current designs lack a principled mechanism for hierarchical bandwidth allocation. In most cases, all transmitted semantic components are treated uniformly, without distinguishing between features that contribute to the global structural context required for scene stability and the high-resolution details necessary for safety-critical long-range detection. This uniform allocation fails to exploit the multi-scale importance distribution inherent in BEV representations, where different feature levels encode complementary spatial and semantic information. Consequently, these methods fail to fully exploit the intrinsic hierarchy of BEV representations to transmit only the minimal sufficient representation and lack robustness in safety-critical scenarios.

Semantic communication has been recently introduced into cooperative perception systems\cite{10405254, 11355867, 10810363} to improve communication efficiency by prioritizing the transmission of task-relevant information. For instances, Sheng et al.\cite{10405254} proposed a semantic cooperative perception framework based on JSCC and an importance map\cite{Where2comm:22} that extracts critical semantic features from LiDAR feature maps before transmission. By selectively transmitting informative semantic components, the framework significantly reduces communication overhead while maintaining perception accuracy under various wireless channel conditions. Based on this idea, Gan et al.\cite{11355867} introduced SComCP, a task-oriented semantic communication framework designed specifically for cooperative perception. In SComCP, an importance-aware feature selection network identifies and prioritizes semantic features that contribute most to the perception objective, enabling selective transmission among cooperating agents. In addition, a JSCC-based semantic codec\cite{8683463} is employed to directly map semantic features into noise-resistant channel symbols, thereby improving robustness against  channel impairments and facilitating reliable communication over noisy and dynamic wireless environments.

Despite these advances, existing semantic communication-based cooperative perception frameworks still face several limitations. First, most current architectures are inherently single-scale, operating on a unified singular feature representation extracted from a specific stage of the perception backbone. Although existing models\cite{10405254, 11355867} effectively prioritize task-relevant regions within this representation, they overlook the hierarchical nature of modern perception networks. This single-scale constraint is fundamentally insufficient for the long-range perception required in high-speed ITS scenarios, as distant actors often manifest as small-scale features that are lost or poorly resolved in a unified feature map, degrading long-range perception reliability. Second, many existing frameworks\cite{11355867} often rely on brute-force symbol redundancy to ensure channel robustness. For instance, projecting a minimal set of semantic features into high-dimensional symbol spaces (e.g., 256 symbols per dimension) inherently limits the volume of transmittable spatial information. This creates a sparsity bottleneck where the system may effectively transmit 12 isolated points, but remains blind to the broader semantic context of the environment, leading to a loss of global scene structure and contextual continuity. 

\section{Methodology}
\subsection{Semantic Communication based Cooperative Perception}
\begin{figure*}[t]
    \centering 
    \captionsetup{justification=centering} % Centers the caption for this figure only
    \includegraphics[width=0.9\textwidth]{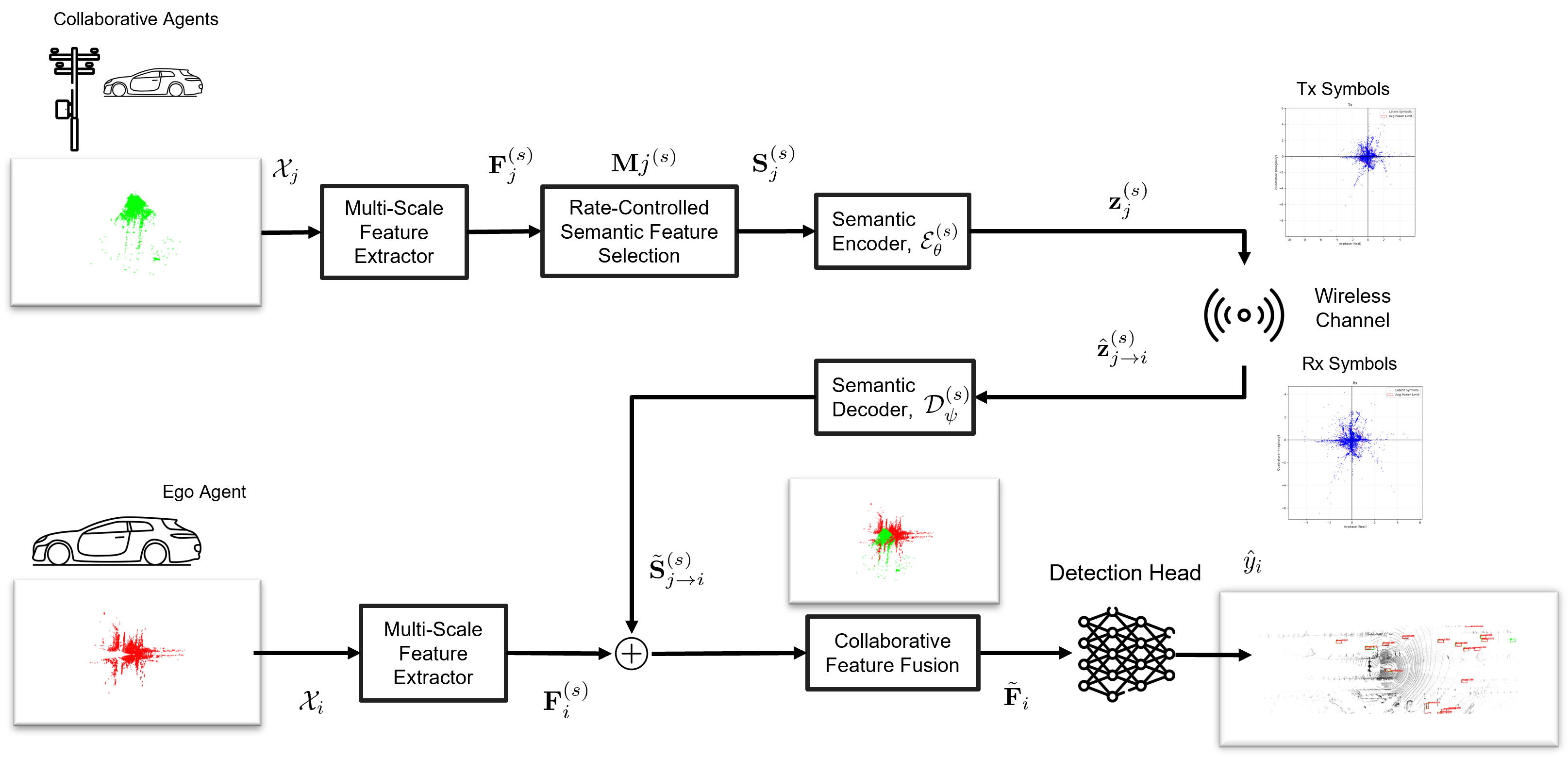}
    \caption{Overview of Hierarchical Multi-Scale Semantic-Aware Cooperative Perception (HMS-SCP) architecture} 
    \label{fig:semcomm} 
\end{figure*}

In this section, the proposed HMS-SCP framework is presented for robust and communication-efficient cooperative perception under practical V2X constraints. First, the task objective is formulated, followed by an overview of the HMS-SCP architecture, as illustrated in Fig.~\ref{fig:semcomm}. In this architecture, multi-scale backbone features are selectively encoded and transmitted through a semantic communication pipeline. In particular, the proposed method introduces a rate-controlled semantic feature selection mechanism that identifies task-critical spatial regions across different scales. These selected features are then processed by a JSCC-based semantic codec, which directly maps them into channel symbols to ensure robustness against wireless fading and noise. Next, the noise-resilient architectural design is discussed with particular emphasis on its semantic codec and the integration with rate-controlled semantic feature selection. Finally, a task-oriented training strategy is presented to align semantic transmission with perception performance, balancing communication efficiency and detection accuracy. 

\subsection{Cooperative Perception Problem Formulation}
Cooperative perception with $N$ agents is considered which may include connected automated vehicles (CAVs) and intelligent infrastructure units under bandwidth and wireless channel constraints. The ego-agent $i$ observes the surrounding environment and cooperates with neighboring agents $j \in \{1, \dots, N\}, j \neq i$. to perform a perception task. Let $X_i$ and $y_i$ denote the observation and corresponding ground-truth label of ego-agent $i$, respectively. Rather than transmitting raw observations, each neighboring agent $j$ extracts a set of hierarchical feature maps from its perception backbone. $\{\mathbf{F}_j^{(s)}\}_{s=1}^{S}$ denote the set of feature maps extracted from $S$ hierarchical levels of the perception backbone. 

To satisfy bandwidth constraints, each agent employs a spatial importance predictor $\mathcal{M}^{(s)}(\cdot)$ to identify task-relevant grid elements at each scale. The predictor generates a binary importance mask $\mathbf{M}_j^{(s)} \in \{0,1\}^{H \times W}$, which identifies the top-$k$ most salient spatial indices at scale $s$. The sparsified semantic representation is then formulated as:
\begin{equation}
\mathbf{S}_j^{(s)} = \mathbf{M}_j^{(s)} \odot \mathbf{F}_j^{(s)},\label{eq:masked_out}
\end{equation}
where $\odot$ denotes the element-wise product with broadcasting across the channel dimension $C$. The resulting sparse features are then mapped to the latent space via the semantic encoder:
\begin{equation}
\mathbf{z}_j^{(s)}= \mathcal{E}_{\theta}^{(s)}(\mathbf{S}_j^{(s)}),\label{eq:encoder}
\end{equation}
where $\mathcal{E}_{\theta}^{(s)}(\cdot)$ denotes the semantic encoder at scale $s$, and $\mathbf{S}_j^{(s)}$ represents the semantic feature at scale $s$. The multi-scale representation captures both fine-grained local information and high-level contextual semantics that are important for robust perception. The received signal at agent $i$ is modeled as:
\begin{equation}
\hat{\mathbf{z}}_{j \to i}^{(s)} = \mathcal{W} \left( \mathbf{z}_j^{(s)}; \gamma \right),\label{eq:wirelesschannel}
\end{equation}
where $\mathcal{W}(\cdot; \gamma)$ represents the wireless channel transfer function (e.g., AWGN or Rayleigh fading) parameterized by the instantaneous SNR $\gamma$. Upon reception, the ego-agent reconstructs the corresponding feature maps using a semantic decoder:
\begin{equation}
\tilde{\mathbf{S}}_{j \to i}^{(s)} = \mathcal{D}_{\psi}^{(s)} \left( \hat{\mathbf{z}}_{j \to i}^{(s)} \right),\label{eq:decoder}
\end{equation}

The reconstructed features are subsequently fused with the local features of ego-agent through a collaborative perception network $\Psi_\Theta(\cdot)$ to produce the final perception output: 
\begin{equation}
\hat{y}_i =
\Psi_\Theta
\left(
\{\mathbf{F}_i^{(s)}\}_{s=1}^{S},
\{\tilde{\mathbf{S}}_{j \to i}^{(s)}\}_{j \ne i,\, s=1}^{S}
\right),
\end{equation}
where $\hat{y}_i$ denotes the predicted perception output for agent $i$,
$\{\mathbf{F}_i^{(s)}\}$ represents the ego-agent's local multi-scale features, and $\{\tilde{\mathbf{S}}_{j \rightarrow i}^{(s)}\}$ denotes the reconstructed multi-scale features received from collaborating agents.

To ensure that the collaborative perception system remains viable under limited wireless bandwidth, the semantic transmission is required to satisfy a global resource constraint
\begin{equation}
\sum_{j \ne i} \sum_{s=1}^{S} R_{j \rightarrow i}^{(s)} \le B,\label{eq:channelused}
\end{equation}
where
\begin{equation}
R_{j \rightarrow i}^{(s)} =
\frac{n_{j \rightarrow i}^{(s)}}{d_{j}^{(s)}},\label{eq:rate}
\end{equation}
denotes the transmission ratio or spatial sampling ratio at scale $s$. Here, $n_{j \rightarrow i}^{(s)}$ corresponds to the number of channel uses, while $d_{j}^{(s)}$ represents the dimensionality of the semantic representation, specifically the spatial grid size.  The transmission ratio $R$ serves as a tunable parameter that enables the framework to dynamically adapt to varying levels of channel congestion characterized by the available bandwidth $B$.

The system is trained in an end-to-end manner to maximize perception performance under varying wireless channel conditions:
\begin{equation}
\max_{\theta,\psi,\Theta}
\sum_{i=1}^{N}
\mathbb{E}_{\gamma \sim p(\gamma)}
\Big[g\big(\hat{y}_i, y_i\big)\Big],
\end{equation}
subject to \eqref{eq:channelused}
where $g(\cdot)$ denotes the evaluation metric of the perception task, and the expectation over $p(\gamma)$ enforces robustness to stochastic channel variation. In this context, the evaluation is done using average precision (AP), which is a commonly used metric to evaluate 3D object detection performance\cite{Geiger2012AreWR}. The multi-scale nature of the framework further allows for dynamic bit-allocation, where finer scales handling nearby objects and coarser scales providing global context for distant objects are assigned different channel uses $n_{j \to i}^{(s)}$ based on their contribution to the perception metric $g(\cdot)$. This allows the framework to dynamically balance communication efficiency and perception accuracy according to their contribution to the overall task objective $g(\cdot)$.

\subsection{HMS-SCP Framework} 

\begin{figure}[!t]
    \centering
    % Using columnwidth is safer for single-column figures
    \includegraphics[width=0.85\columnwidth]{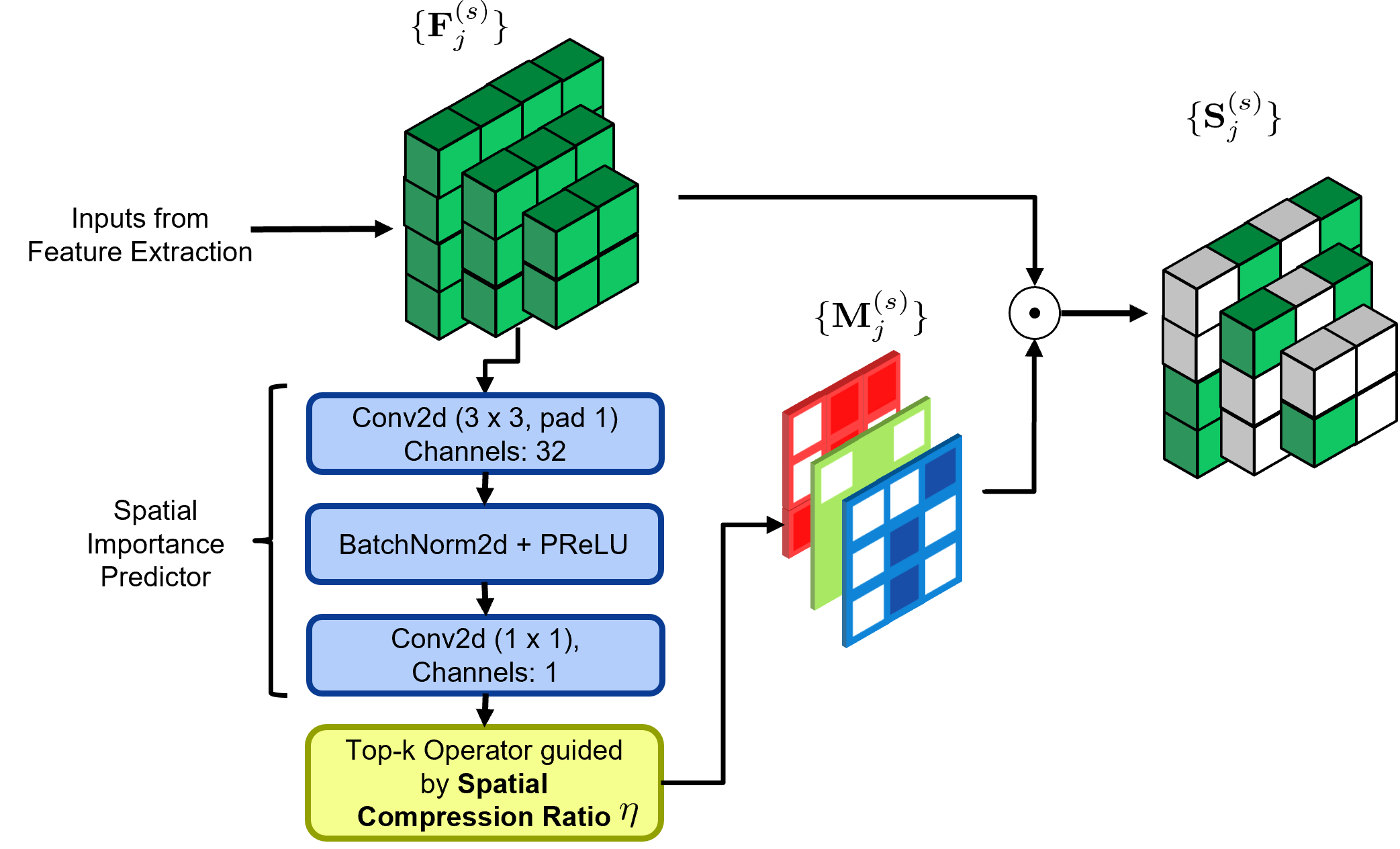}
    \caption{Rate-controlled semantic feature selection process.}
    \label{fig:sparsity}
\end{figure}
\begin{figure*}[htbp]
    \centering 
    \captionsetup{justification=centering} 
    % Added height constraint and keepaspectratio to prevent the "too large" error
    \includegraphics[width=0.85\textwidth, height=0.85\textheight, keepaspectratio]{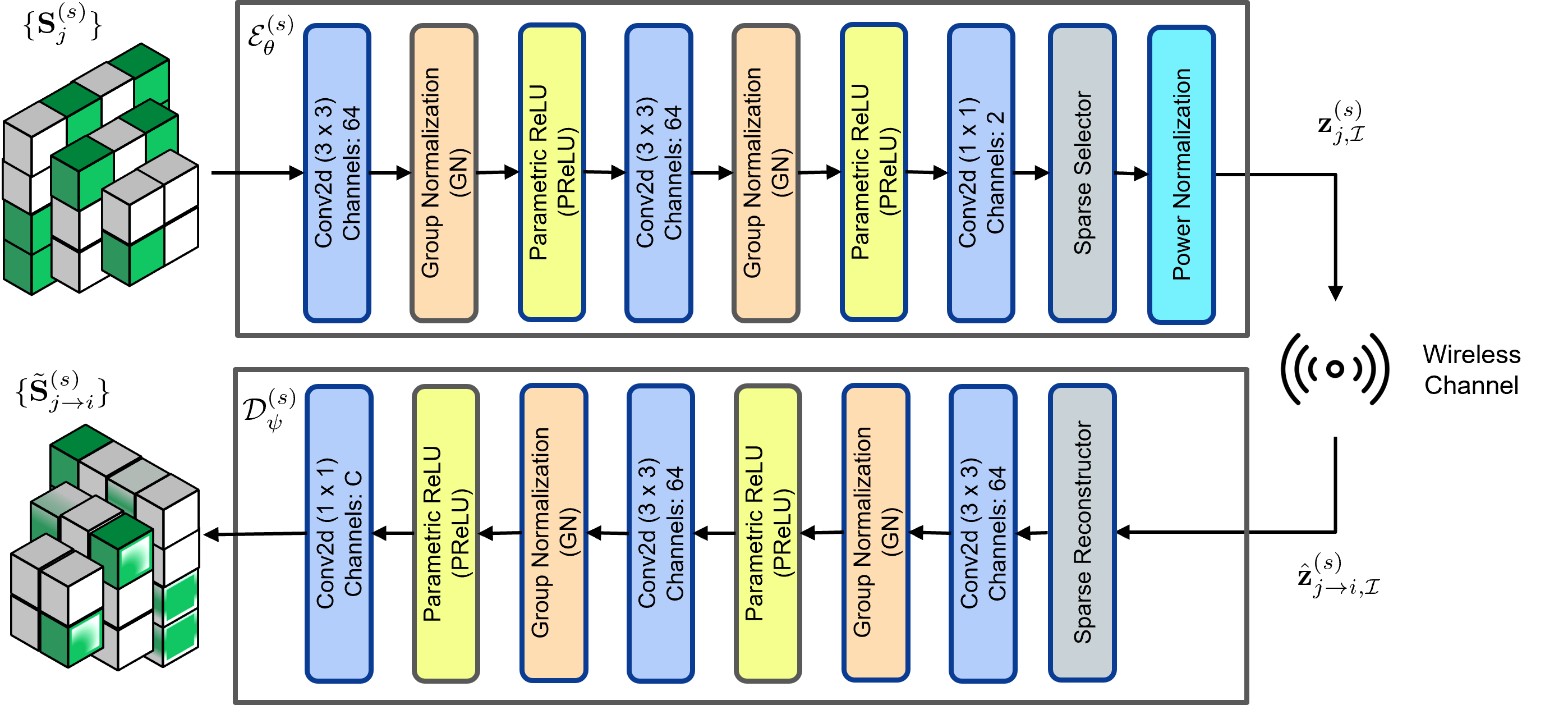}
    \caption{The architecture of our proposed semantic codec} 
    \label{fig:codec} 
\end{figure*}
The overall architecture of the proposed HMS-SCP framework as illustrated in Fig.~\ref{fig:semcomm}, consists of five major components: multi-scale feature extraction, rate-controlled semantic feature selection, semantic codec, collaborative feature fusion and object detection. The transmission of ego agent's metadata, including poses and extrinsics, is assumed to be well synchronized, providing the necessary spatial information for all collaborative agents to project their LiDAR point clouds into the coordinate frame of the ego-agent prior to feature extraction. In practical C-ITS deployments, such spatial metadata can be reliably disseminated through standardized protocols such as the Collective Perception Message (CPM)\cite{ETSI_TS_103_324_V2.1.1, 11077784} defined in the ETSI Collective Perception Service (CPS), facilitating seamless multi-agent coordination.

\textbf{Multi-Scale Feature Extraction:} The feature extraction module processes LiDAR point cloud data as the primary sensory input for cooperative perception. For each neighboring agent $j$, the raw point cloud $\mathcal{X}_j$ is first encoded using a Pillar Feature Encoding (PFE) layer \cite{8954311}, which transforms sparse 3D points into a structured pillar-based representation. Through a pillar scattering operation, the encoded features are projected onto a 2D BEV pseudo-image, yielding a spatial representation map $\mathbf{F}_j \in \mathbb{R}^{H \times W \times C}$. The resulting BEV feature map is then processed by a hierarchical convolutional backbone to extract spatial features at multiple resolutions\cite{10160546}. Specifically, the backbone progressively downsamples the feature maps through a stacked of convolutional layers, producing a set of multi-scale feature representations 
$\{\mathbf{F}_j^{(1)}, \mathbf{F}_j^{(2)}, \mathbf{F}_j^{(3)}\}$ corresponding to $2\times$, $4\times$, and $8\times$ downsampled resolutions. These hierarchical features provide complementary information across different receptive fields, encoding both fine-grained spatial details and high-level contextual semantics, and serve as inputs to the subsequent rate-controlled semantic feature selection module.

\textbf{Rate-Controlled Semantic Feature Selection:}
To satisfy communication constraints, each agent $j$ employs a learnable spatial importance predictor to estimate the contribution of each BEV location to the downstream perception task. For each scale $s$, the module takes the feature map $\mathbf{F}_j^{(s)}$ as input to generate an importance logit map, which is transformed via a sigmoid activation to produce a probabilistic saliency map:
\begin{equation}
\mathbf{P}_j^{(s)} \in (0,1)^{H_s \times W_s}.
\end{equation} This map represents the pixel-wise probability of a spatial location that contributes to the perception task. As shown in Fig.~\ref{fig:sparsity}, a spatial compression ratio $\eta \in (0, 1)$ is then introduced to directly govern the transmission budget. By selecting the top-$k$ spatial indices corresponding to the $k$ largest values in the saliency map $\mathbf{P}_j^{(s)}$, a fixed sparsity constraint is enforced as:
\begin{equation}
k = \lfloor \eta \cdot H_s W_s\rceil. \label{eq:ksel} 
\end{equation}
This operation yields a binary selection mask $\mathbf{M}_j^{(s)} \in \{0,1\}^{H_s \times W_s}$, where only the most informative spatial indices corresponding to the $k$ highest-scoring positions in $\mathbf{P}_j^{(s)}$ are set to 1. In the forward pass, the importance predictor $\mathcal{M}^{(s)}(\cdot)$ applies this discrete mask to the feature maps $\mathbf{F}_j^{(s)}$, resulting in the masked multi-scale semantic representation $\mathbf{S}_j^{(s)}$ defined in \eqref{eq:masked_out}. The physical mapping of this representation using spatial compression ratio $\eta$ to directly dictate the communication overhead is further elaborated in the subsequent subsection.

\textbf{Semantic Codec:} As shown in Fig.~\ref{fig:codec}, the semantic codec pipeline is designed to compress and reconstruct the masked multi-scale semantic features $\mathbf{S}_j^{(s)} \in\mathbb{R}^{H_s \times W_s \times C_s}$ to enable communication-efficient feature. For each scale $s \in \{1, \dots, S\}$, the feature map is transformed by a scale-specific semantic encoder $\mathcal{E}_{\theta}^{(s)}$ to produce a compact latent representation $\mathbf{z}_j^{(s)}$ as defined in \eqref{eq:encoder}. This scale-specific encoding design allows each hierarchical feature level to be compressed according to its own spatial resolution and semantic characteristics, rather than forcing all features into a unified single-scale representation. The dimension of $\mathbf{z}_j^{(s)}$ determines the number of symbols for transmission. The encoder $\mathcal{E}_{\theta}^{(s)}$ utilizes a sequential convolutional architecture designed to compress spatial redundancy while preserving critical semantic information. Under JSCC, the encoded latent representation is directly mapped into channel symbols and transmitted over a wireless channel, resulting in the received signal $\hat{\mathbf{z}}_{j \to i}^{(s)}$ as defined in \eqref{eq:wirelesschannel}. At the receiver, the ego-agent employs a semantic decoder $\mathcal{D}_{\psi}^{(s)}$ to reconstruct the feature maps $\tilde{\mathbf{S}}_{j \to i}^{(s)}$ as defined in \eqref{eq:decoder}. The detailed noise-resilient architectural design of the codec is further discussed in the subsequent subsection.

\begin{figure}[!t]
\centering
\includegraphics[width=3.5in]{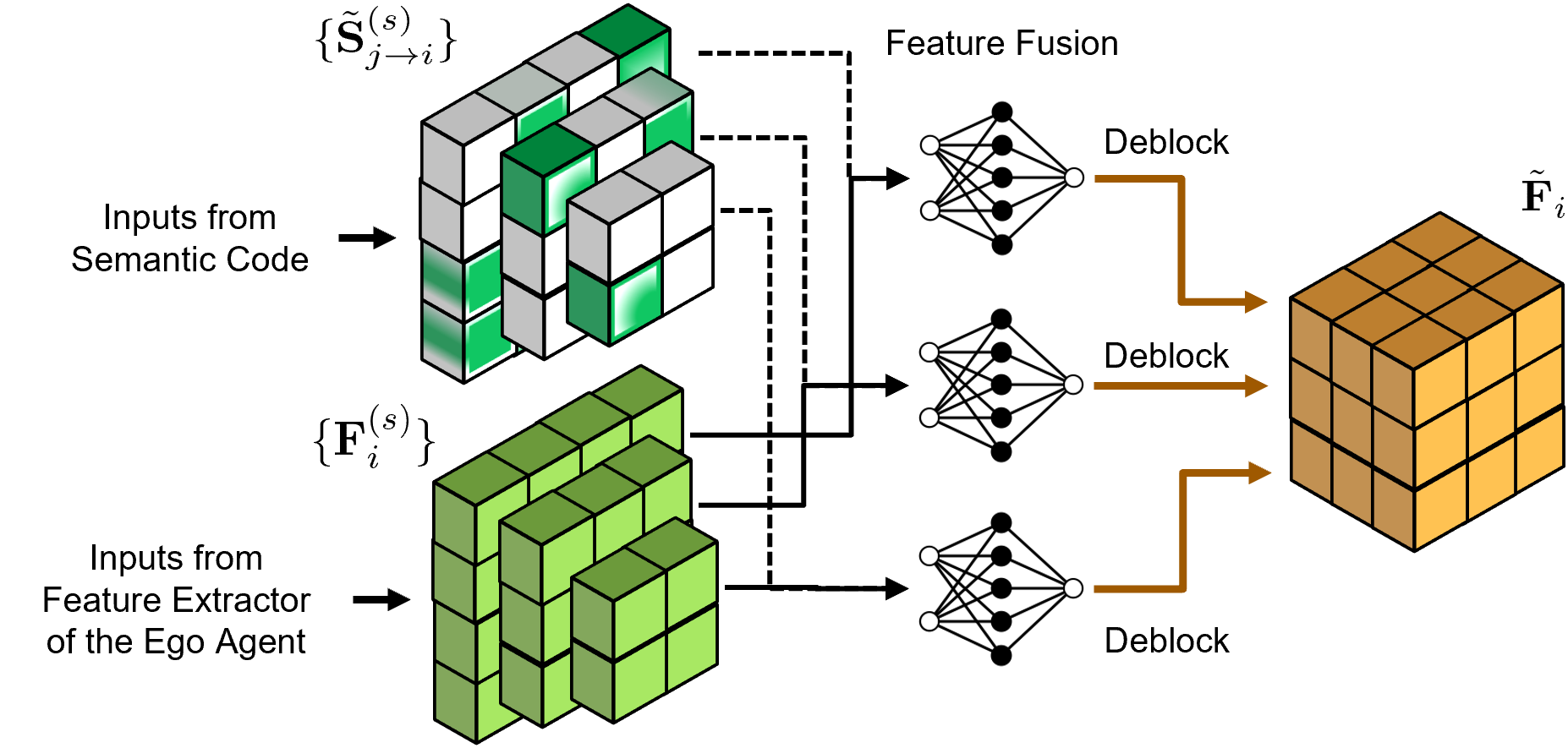}
\caption{Collaborative feature fusion process.}
\label{fig:fusion}
\end{figure}

\textbf{Collaborative Feature Fusion:} As illustrated in Fig.~\ref{fig:fusion}, the ego-agent $i$ aggregates the reconstructed multi-scale semantic representations $\{\tilde{\mathbf{S}}_{j \to i}^{(s)}\}$ received from all neighboring agents $j \in \{1, \dots, N\}, j \neq i$. In contrast to single-scale methods, the fusion process is performed independently for each of the $S$ hierarchical scales, thereby preserving scale-specific spatial and semantic characteristics throughout the aggregation process. For each scale $s$, a dedicated fusion block is employed to perform the following operations: \textbf{Spatial Alignment}, which applies a warp affine transformation derived from pairwise spatial metadata to align neighbor features with the ego agent's BEV coordinate system; and \textbf{Attentional Aggregation}, which employs a a Scaled Dot-Product Attention mechanism\cite{10.5555/3295222.3295349} to fuse the aligned features from all neighboring agents with the ego agent's own feature representation $\mathbf{F}_i^{(s)}$. For each spatial location, the module computes a weighted sum of agent-wise features as follows:
\begin{equation}
    \text{Attention}(\mathbf{Q}, \mathbf{K}, \mathbf{V}) = \text{softmax}\left(\frac{\mathbf{Q}\mathbf{K}^\top}{\sqrt{d_k}}\right)\mathbf{V},
    \end{equation}
where attention is performed across the agent dimension. This operation transforms multiple per-agent feature tensors into a unified ego-centric scene representation, enabling the model to emphasize reliable complementary observations while suppressing features corrupted by occlusion, misalignment, or channel-induced distortion.

Importantly, the spatial dimensions and channel depths of each scale are preserved during this stage. Following per-scale fusion, a multiscale decoding module upsamples each fused representation to a common high-resolution spatial grid. Specifically, the fused features are processed by scale-specific deconvolutional blocks resulting in a unified set of feature maps and then concatenated along the channel dimension. Finally, the channel dimension is further reduced to produce the final comprehensive BEV representation $\tilde{\mathbf{F}}_i \in \mathbb{R}^{H \times W \times C}$, where $C$ represents the channel dimension set to 256 in our implementation for computational efficiency. This fused volume captures a hierarchical range of geometric and semantic information, serving as the input for the downstream 3D object detection head. 

\textbf{Object Detection and Multi-Task Optimization}: The detection head is optimized using a composite loss $\mathcal{L}_{\text{det}}$, comprising classification, regression, and orientation components. For classification, the \textit{Focal Loss}\cite{8417976} formulation is adopted to mitigate the inherent class imbalance between sparse foreground objects and the dominant background regions. The regression branch supervises the 3D bounding box parameters $(x, y, z, w, l, h, \theta)$, where $(x, y, z)$ denotes the center position, $(w, l, h)$ represents the physical dimensions, and $\theta$ indicates the yaw angle. The \textit{Smooth L1 loss}\cite{7410526} is employed for bounding-box regression, together with an auxiliary direction loss\cite{8954311} to resolve the $180^\circ$ orientation ambiguity.

A multi-scale JSCC reconstruction loss is integrated to preserve the semantic fidelity of communicated features under channel impairments. Since the proposed semantic codec processes representations across $S$ hierarchical resolutions, the reconstruction error is calculated independently at each scale by comparing the masked sender feature map before transmission, $\mathbf{S}_j^{(s)}$, with its reconstructed counterpart at the receiver, $\tilde{\mathbf{S}}_{j \to i}^{(s)}$. The overall reconstruction loss is obtained by averaging the per-scale losses:
\begin{equation}
    \mathcal{L}_{\text{rec}} = \frac{1}{S} \sum_{s=1}^{S} \mathbb{E} \left[ \| \tilde{\mathbf{S}}_{j \to i}^{(s)} - \mathbf{S}_j^{(s)} \|_2^2 \right].\label{eq:recloss}
\end{equation}

The final stage of the pipeline evaluates perception performance by applying a multi-task loss function to the fused BEV feature maps. A joint optimization strategy has been adopted to balance the primary detection objective with semantic reconstruction fidelity across all scales:
\begin{equation}
    \mathcal{L}_{\text{total}} = \mathcal{L}_{\text{det}}(\hat{y}_i, y_i) + \mu \cdot \mathcal{L}_{\text{rec}},\label{eq:totalloss}
\end{equation}
where $\mu$ is a scaling hyperparameter that controls the contribution of the reconstruction term. By minimizing $\mathcal{L}_{\text{rec}}$ alongside the detection loss $\mathcal{L}_{\text{det}}$, the network learns to prioritize the preservation of task-critical semantic information across all hierarchical levels. This ensures that the multi-scale decoder can faithfully reconstruct high-dimensional features from the sparse symbol stream. This joint optimization enables the detection head to leverage both high-resolution geometric details and global context for robust 3D bounding box prediction, even under stringent bandwidth constraints and imperfect wireless channel conditions.

\subsection{Noise-Resilient Architectural Design}

To ensure high-fidelity reconstruction under imperfect channel conditions, a noise-resilient architecture for the semantic codec $(\mathcal{E}_\theta, \mathcal{D}_\psi)$ as illustrated in Fig.~\ref{fig:codec} has been proposed. The encoder $\mathcal{E}_\theta$ utilizes a sequential multi-scale convolutional design, where two stacked $3 \times 3$ are first employed to capture local spatial correlations followed by a $1 \times 1$ bottleneck layer that projects features into the latent space. This stacked design enables the encoder to capture richer local spatial dependencies and perform deeper nonlinear feature transformation prior to the latent projection, which helps produce a more robust semantic representation under channel distortion. 

To further improve reconstruction stability, Group Normalization (GN)\cite{wu2018group} and Parametric ReLU (PReLU) activations\cite{8683463, 9066966} is integrated throughout the encoder-decoder pipeline. Compared with Batch Normalization (BN), GN provides stable feature normalization independent of batch size, which is important in decentralized collaborative perception, where the number of communicating neighbors and transmitted features can vary across scenes. Meanwhile, PReLU preserves negative responses via a learnable slope, adaptively reshaping feature distributions and avoiding information loss from hard zero-clipping. This is particularly beneficial in semantic communication, where weak yet informative features aid robust reconstruction under noise and fading. On the decoder, PReLU retains negative-valued signals to enhance recovered BEV feature fidelity for downstream tasks.

To satisfy the physical constraints of the wireless transmitter, the latent representation $\mathbf{z}_j^{(s)}$ is mapped into a complex-valued signal space. Specifically, the output channel dimension is interpreted as a paired set of real and imaginary components, such that
\begin{equation}
    \mathbf{z}_j^{(s)} = [\mathbf{z}_{\mathrm{re}}, \mathbf{z}_{\mathrm{im}}],
\end{equation}
where $\mathbf{z}_{\mathrm{re}}, \mathbf{z}_{\mathrm{im}} \in \mathbb{R}^{H_s \times W_s \times C{\text{out}}}$, with $C_{\text{out}}$ denoting the number of complex-valued symbols assigned to each spatial location. Unlike conventional dense transmission, our framework performs sparse latent transmission. Following the rate-controlled selection mechanism, only the latent vectors at spatial indices $u \in \mathcal{I}_j^{(s)}$, defined by the mask $\mathbf{M}_j^{(s)}$, are transmitted. By setting $C_{out}=1$, each selected location is mapped to a single complex-valued symbol $\mathbf{z}_{j,u}^{(s)} \in \mathbb{C}$. The resulting serialized sequence $\mathbf{z}_{j,\mathcal{I}}^{(s)} = ( \mathbf{z}_{j,u}^{(s)} )_{u \in \mathcal{I}_j^{(s)}}$ ensures the transmission volume is exactly $K_s = |\mathcal{I}_j^{(s)}| = k$. Consequently, the total channel uses are governed by $n_j = \sum_{s=1}^{S} |\mathbf{M}_j^{(s)}|_0$, and the normalized transmission ratio simplifies to $R = \eta$. This establishes a direct correspondence between the spatial sparsity budget and actual channel usage, enabling precise and interpretable control over communication overhead.

To regulate the average transmit power to a fixed constraint $P$, a power normalization operation $\phi(\cdot)$ is applied  to the sparse latent sequence prior to channel perturbation. The normalized transmit symbols are expressed as
\begin{equation}
    \mathbf{z}_{j \to i,\mathcal{I}}^{(s)} =
    \mathbf{z}_{j,\mathcal{I}}^{(s)}
    \cdot
    \sqrt{
       \frac{P}{
         \frac{1}{K_s}
         \sum_{u \in \mathcal{I}_j^{(s)}}
         \left\|
         \mathbf{z}_{j,u}^{(s)}
         \right\|^2
         + \epsilon
         }
    },
\end{equation}
where $\epsilon$ is a small constant introduced for numerical stability. To prevent gradient explosion particularly during the initial stages of training, a scale clipping mechanism that upper-bounds the normalization factor at a maximum value $\lambda_{\max}$ is employed. This prevents the transmission of excessively high-energy symbols when the encoder produces latent representations with small magnitudes, which is particularly important during early training or under highly sparse feature activation. The normalized sparse symbols are then transmitted over a wireless channel. Let \(\mathbf{z}_{j \to i,\mathcal{I}}^{(s)}\) denote the symbols received at agent \(i\). The channel input-output relationship is modeled as:
\begin{equation}
\hat{\mathbf{z}}_{j \to i, \mathcal{I}}^{(s)} = h_{j \to i}^{(s)} \mathbf{z}_{j \to i, \mathcal{I}}^{(s)} + \mathbf{n}_{j \to i, \mathcal{I}}^{(s)},
\end{equation}
where \(h_{j \to i}^{(s)} \in \mathbb{C}\) is the channel coefficient and \(\mathbf{n}_{j \to i}^{(s)} \sim \mathcal{CN}(\mathbf{0}, \sigma_n^2 \mathbf{I})\) denotes additive circularly symmetric complex Gaussian noise. For the AWGN setting, \(h_{j \to i}^{(s)}\) is set to 1, such that the received signal is corrupted by additive noise only. For the Rayleigh fading setting, \(h_{j \to i}^{(s)}\) is sampled from a zero-mean complex Gaussian distribution, i.e., 
\begin{equation}
h_{j \to i}^{(s)} \sim \mathcal{CN}(0,1),
\end{equation}
which models small-scale multipath fading with Rayleigh-distributed amplitude. Specifically, the effective average signal-to-noise ratio for both channel settings is defined as:
\begin{equation}
\gamma = \frac{P}{\sigma_n^2}
\end{equation}
where $P$ is the average power constraint per complex-valued symbol (set to $1$ in our implementation without loss of generality) and $\sigma_n^2$ denotes the noise variance.

After transmission, the received sparse symbols $\hat{\mathbf{z}}_{j \to i, \mathcal{I}}^{(s)}$ are re-mapped to their original spatial coordinates based on the index set $\mathcal{I}_j^{(s)}$. Locations not selected for transmission are zero-padded, forming a dense latent feature map for decoding. The decoder $\mathcal{D}_\psi$, which mirrors the encoder structure, progressively upsamples these symbols through a symmetric set of GN-stabilized convolutional blocks to restore the feature map $\tilde{\mathbf{S}}_{j\to i}^{(s)}$. By modeling the system as an end-to-end differentiable pipeline, the proposed framework abstracts the packet-level protocols while strictly adhering to the power and bandwidth constraints. This formulation yields semantic representations inherently resilient to stochastic wireless impairments, enabling robust cooperative perception in safety-critical V2X environments.

\subsection{Task-Oriented Training Strategy}
\begin{algorithm}[t]
\small % Use small font to ensure it fits on one page
\caption{Two-Stage Training Strategy}
\label{alg:training_strategy}
\SetKwInOut{Input}{Input}
\SetKwInOut{Variables}{Let}

\Input{Dataset $\mathcal{D}=\{X_j, y_j\}_{j=1}^N$, Spatial compression ratio $[\eta_{\min}, \eta_{\max}]$, SNR $[\gamma_{\min}, \gamma_{\max}]$}
\Variables{$\mathcal{E}_\theta, \mathcal{D}_\psi$: Codec; $\mathcal{W}$: Wireless Channel; $\Psi_\Theta$: Collaborative Perception Network; $\mathcal{M}^{(s)}$: Spatial Importance Predictor.}

\BlankLine
\textbf{Stage 1: Feature Reconstruction}\;
\While{not converged}{
    Sample $\mathcal{B} \subset \mathcal{D}$\;
    \ForEach{$(X_j, y_j) \in \mathcal{B}$}{
        $\{\mathbf{F}_j^{(s)}\} \leftarrow \text{Backbone}(X_j)$\;
        \For{$s=1$ \KwTo $S$}{
            $\mathbf{M}_j^{(s)} \leftarrow \text{Top-}k(\mathcal{M}^{(s)}(\mathbf{F}_j^{(s)}), \eta)$\;
            $\hat{\mathbf{z}}_{j \to i}^{(s)} \leftarrow \mathcal{W}(\text{Norm}(\mathcal{E}_\theta^{(s)}(\mathbf{F}_j^{(s)} \odot \mathbf{M}_j^{(s)})); \gamma)$\;
            $\tilde{\mathbf{S}}_{j \to i}^{(s)} \leftarrow \mathcal{D}_\psi^{(s)}(\hat{\mathbf{z}}_{j \to i}^{(s)})$\;
        }
    }
    Compute reconstruction loss: $\mathcal{L}_{\text{rec}}$, defined as \eqref{eq:recloss}\;
    Update $\{\theta, \psi, \mathcal{M}\}$ via STE to minimize $\mathcal{L}_{\text{rec}}$\;
}

\BlankLine
\textbf{Stage 2: Task-Oriented Joint Optimization}\;
\textbf{Freeze:} Backbone\; 
\textbf{Train:} $\{\theta, \psi, M, \Theta\}$\;
\While{not converged}{
    Sample $\mathcal{B} \subset \mathcal{D}$\;
    \ForEach{ego $i$ and collaborators $j \in \mathcal{B}$}{
            \For{$s = 1$ \KwTo $S$}{
                $\mathbf{M}_j^{(s)} \leftarrow \text{Top-k}(\mathcal{M}^{(s)}(\mathbf{F}_j^{(s)}), \eta)$\;
                $\hat{\mathbf{z}}_{j \to i}^{(s)} \leftarrow \mathcal{W}(\text{Norm}(\mathcal{E}_{\theta}^{(s)}(\mathbf{F}_j^{(s)} \odot \mathbf{M}_j^{(s)})); \gamma)$\;
                $\tilde{\mathbf{S}}_{j \to i}^{(s)} \leftarrow \mathcal{D}_{\psi}^{(s)}(\hat{\mathbf{z}}_{j \to i}^{(s)})$\;
            }
            $\hat{y}_i \leftarrow \Psi_\Theta \left( \{\mathbf{F}_i^{(s)}\}_{s=1}^S, \{\tilde{\mathbf{S}}_{j \to i}^{(s)}\}_{j \neq i, s=1}^S \right)$\;
    }
    Compute total loss: $\mathcal{L}_{\text{total}}$, defined as \eqref{eq:totalloss}\;
    Jointly update $\{\theta, \psi, \mathcal{M}, \Theta\}$ to minimize $\mathcal{L}_{\text{total}}$\;
}

\end{algorithm}
As detailed in Algorithm~\ref{alg:training_strategy}, a two-stage optimization strategy is adopted to ensure stable convergence and align semantic transmission with the downstream perception task. Training is initialized from a pre-trained model without semantic codec and spatial importance predictor. In the first stage, the semantic codec ($\mathcal{E}_\theta, \mathcal{D}_\psi$) and the spatial importance predictor are jointly trained to minimize the multi-scale JSCC reconstruction loss as defined in \eqref{eq:recloss}. This stage aims to establish a reliable semantic reconstruction capability before task-level fine-tuning. To improve robustness under realistic non-LoS V2X conditions, the training process is conducted over a wide range of Rayleigh fading channels with the SNR $\gamma$ sampled from $\mathcal{U}(0, 20)$~dB. In parallel, the spatial compression ratio $\eta$ is randomly sampled from $\mathcal{U}(0.01, 0.20)$, allowing the model to learn stable feature selection and reconstruction across different bandwidth budgets. Specifically, the Straight-Through Estimator (STE) technique is employed to enable end-to-end backpropagation through the non-differentiable Top-$k$ selection mask, allowing the importance predictor to learn a task-relevant ranking of spatial features.

In the second stage, the optimization objective shifts from signal-level reconstruction to task-oriented perception accuracy. The feature extraction backbone initialized from the pre-trained model is frozen to preserve learned geometric representation and prevent the catastrophic forgetting of geometric features. The joint fine-tuning of the remaining architecture, encompassing the hierarchical multi-scale semantic codec, the spatial importance predictor, the downstream fusion and the detection head, is performed. By minimizing the composite loss function $\mathcal{L}_{\text{total}}$ in \eqref{eq:totalloss}, the framework learns to prioritize semantic primitives that are most critical for 3D object detection under constrained communication. This selective optimization allows the fusion and neck layers to adaptively compensate for channel-induced distortions, ensuring that the synthesized BEV representation $\tilde{\mathbf{F}}_i$ is optimally conditioned for perception accuracy rather than merely structural similarity. Specifically, Rayleigh fading channels and spatial compression ratios in this training stage are incorporated using a methodology consistent with the first stage.

\section{Experiments}
\subsection{Experimental Setup}

\textbf{Datasets:} The proposed HMS-SCP framework is evaluated on two different cooperative perception datasets: the simulated OPV2V\cite{9812038} and real-world DAIR-V2X\cite{9879243} datasets covering both V2V and V2I collaboration respectively. The evaluation on the OPV2V dataset is conducted within a spatial range of $x \in [-140.8, 140.8]$ m and $y \in [-38.4, 38.4]$~m with a maximum collaboration radius of 70~m. The HMS-SCP framework is also further evaluated on the DAIR-V2X dataset within a spatial range of $x \in [-100.8, 100.8]$~m and $y \in [-40, 40]$~m with a maximum collaboration radius of 100~m. 
\begin{table}[t]
\centering
\caption{Comparison of different semantic-aware cooperative perception frameworks.}
\label{tab:comparison}
\setlength{\tabcolsep}{3pt}
\renewcommand{\arraystretch}{1.1}

\begin{tabular}{lccccc}
\toprule
\textbf{Method} & \textbf{Ch.} & \textbf{Imp.} & \textbf{JSCC} & \textbf{Sym.} & \textbf{Scale} \\
\midrule
Where2Comm (SS) & $\times$ & $\checkmark$ & $\times$ & -- & SS \\
Where2Comm (MS) & $\times$ & $\checkmark$ & $\times$ & -- & MS \\
SComCP          & $\checkmark$ & $\checkmark$ & $\checkmark$ & 256 & SS \\
HMS-SCP (SS)       & $\checkmark$ & $\checkmark$ & $\checkmark$ & \textbf{1} & SS \\
HMS-SCP       & $\checkmark$ & $\checkmark$ & $\checkmark$ & \textbf{1} & MS \\
\bottomrule
\end{tabular}

\vspace{2pt}
{\footnotesize \emph{Ch.: Channel-resilient; Imp.: Importance predictor; Sym.: Complex-valued symbols per selected feature dimension; SS: Single-scale; MS: Multi-scale.}}

\end{table}

\textbf{Implementation:} The HMS-SCP framework is implemented based on the OpenCOOD\cite{9812038} framework, using PointPillar\cite{8954311} adopted as the LiDAR feature encoder. The voxel grid size is set to (0.4~m, 0.4~m), and a maximum of 32 points per Pillar. The number of hierarchical scales, $S$ is set to 3 to provide a balanced multi-resolution representation for cooperative perception. The finest scale preserves high-resolution geometric details that are important for detecting small or distant objects, the intermediate scale captures object-level semantic structures, and the coarsest scale encodes broader contextual information for scene-level reasoning. Adding more scales incurs higher complexity in semantic coding, feature fusion, and bandwidth scheduling, while excessively coarser features offer marginal perceptual gains. Therefore, $S = 3$ offers an effective trade-off between semantic richness, communication efficiency, and real-time inference feasibility. 

To evaluate performance under different communication budgets, the bandwidth constraints are configured by setting a target transmission ratio of $R \in \{0.01, 0.05, 0.10\}$. This corresponds to an extreme $100\times$, significant $20\times$, and moderate $10\times$ spatial reduction regime, respectively. For each configuration, the importance predictor is calibrated to select the top $k$ percentile of spatial features according to the prescribed transmission ratio, while maintaining the single complex-valued symbol bottleneck per feature. To emulate practical wireless environments, the model is trained under Rayleigh fading channels with the SNR uniformly sampled from $[0, 20]$ dB, and evaluated under both AWGN and Rayleigh fading conditions. In addition, the spatial compression ratio $\eta$ is dynamically sampled from $[0.01, 0.20]$ during both training stages to improve robustness across varying communication budgets. This strategy encourages the importance predictor to learn a generalized spatial ranking rather than overfitting to a fixed transmission ratio.

The Adam optimizer is employed for training, with learning rates set to 0.0002 for stage 1 and 0.0001 for stage 2. The loss balancing coefficient in \eqref{eq:totalloss} is set to $\mu = 0.05$ for stage 2, prioritizing task-specific detection loss $\mathcal{L}_{\text{det}}$ while using reconstruction loss $\mathcal{L}_{\text{rec}}$ as a mild regularizer. This configuration preserves physically meaningful latent representations without sacrificing the precision required for 3D object detection under low-SNR conditions. Unless otherwise specified, all remaining experimental settings follow the OpenCOOD configuration. All models are trained and evaluated on a workstation equipped with an NVIDIA RTX A5000 GPUs and an Intel i9-13900 CPU, with latency measured on the same platform.

\subsection{Comparative Performance Analysis}

In this context, AP@0.5 and AP@0.7 are adopted as the primary evaluation metrics. The proposed HMS-SCP framework is compared against two representative SOTA cooperative perception methods, namely Where2Comm\cite{Where2comm:22} and SComCP\cite{11355867}. As the official implementation of SComCP is not publicly available, a baseline implementation is developed following the architectural descriptions and specifications reported in the original work. This includes the use of PointPillar encoder for feature extraction, importance-aware feature selection, single-scale attention-based fusion and a projection to 256 symbols per feature. Previous studies\cite{11355867} indicate that Where2Comm exhibits performance degradation when the importance map is enabled. Consequently, its compression mechanism is disabled in the main comparison to assess its baseline perception capability, as Where2Comm is not inherently designed for channel-resilient transmission. In addition, a multi-scale variant of Where2Comm is implemented to examine whether hierarchical fusion alone can improve robustness beyond the single-scale configurations commonly considered in prior semantic communication studies\cite{10405254, 11355867}. To further isolate the contribution of the proposed hierarchical semantic communication design, an additional single-scale (SS) variant of HMS-SCP is evaluated. In this configuration, the multi-scale features $\mathbf{F}_j^{(s)}$ are first passed through deconvolutional blocks and then concatenated at the transmitter to form a unified feature maps $\mathbf{S}_j$. A single importance mask is generated for this concatenated volume, and the selected features are transmitted in a single pass. Finally, the attention-based fusion mechanism is applied to the reconstructed single-scale feature map $\tilde{\mathbf{S}}_{j \to i}$. Table~\ref{tab:comparison} compares our framework with existing baselines.
\begin{table}[h] 
\centering
\caption{Performance Gap (\%) of $\text{HMS-SCP}$ \\Relative to the Upper Bound at $\text{SNR} = 0\,\mathrm{dB}$.}
\label{table:performance_gap_ms}
\renewcommand{\arraystretch}{1.2}
\resizebox{\columnwidth}{!}{%
\begin{tabular}{@{}llcccc@{}}
\toprule
\textbf{Dataset} & \textbf{Channel} & \textbf{Metric} & \textbf{$R=0.01$ $\downarrow$} & \textbf{$R=0.05$ $\downarrow$} & \textbf{$R=0.10$ $\downarrow$} \\ \midrule
\multirow{4}{*}{\textbf{OPV2V}} & \multirow{2}{*}{Rayleigh} & AP@0.5 & 1.88 & 0.71 & 0.62 \\
 &  & AP@0.7 & 2.92 & 1.11 & 1.17 \\ \cmidrule(l){2-6} 
 & \multirow{2}{*}{AWGN} & AP@0.5 & 1.67 & 0.51 & 0.36 \\
 &  & AP@0.7 & 2.55 & 0.65 & 0.45 \\ \midrule
\multirow{4}{*}{\textbf{DAIR-V2X}} & \multirow{2}{*}{Rayleigh} & AP@0.5 & 1.99 & 0.75 & 0.73 \\
 &  & AP@0.7 & 1.76 & 0.35 & 0.35 \\ \cmidrule(l){2-6} 
 & \multirow{2}{*}{AWGN} & AP@0.5 & 1.84 & 0.58 & 0.55 \\
 &  & AP@0.7 & 1.40 & 0.22 & 0.05 \\ \bottomrule
\end{tabular}%
}
\end{table}

\begin{figure*}[!t]
    \centering
    \captionsetup[subfloat]{font=scriptsize}
    
    % Row 1: OPV2V AP@0.5 & AP@0.7
    \subfloat[Detection AP@0.5 (AWGN)]{\includegraphics[width=0.24\textwidth]{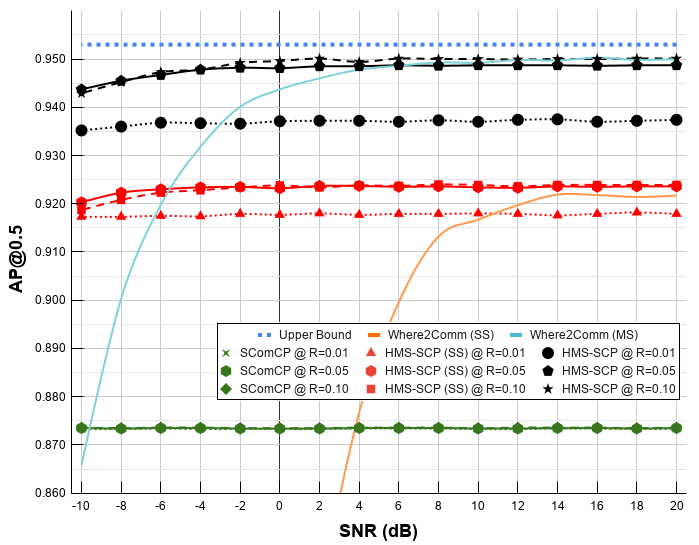}\label{fig:1a}}
    \hfill
    \subfloat[Detection AP@0.5 (Rayleigh)]{\includegraphics[width=0.24\textwidth]{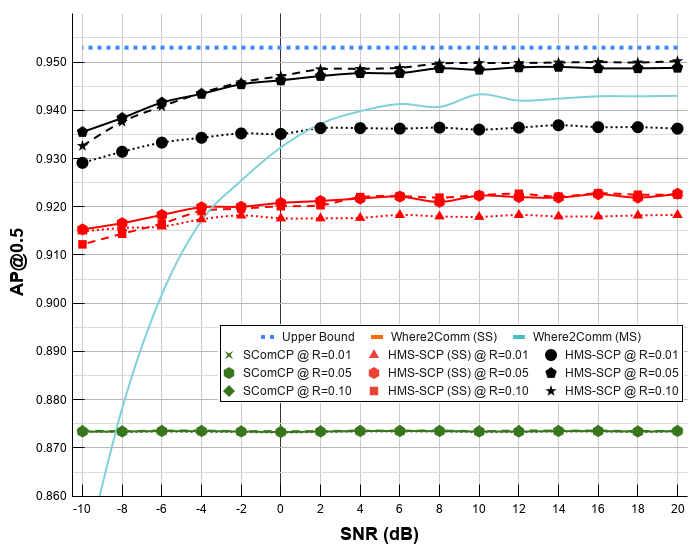}\label{fig:1b}}
    \hfill
    \subfloat[Detection AP@0.7 (AWGN)]{\includegraphics[width=0.24\textwidth]{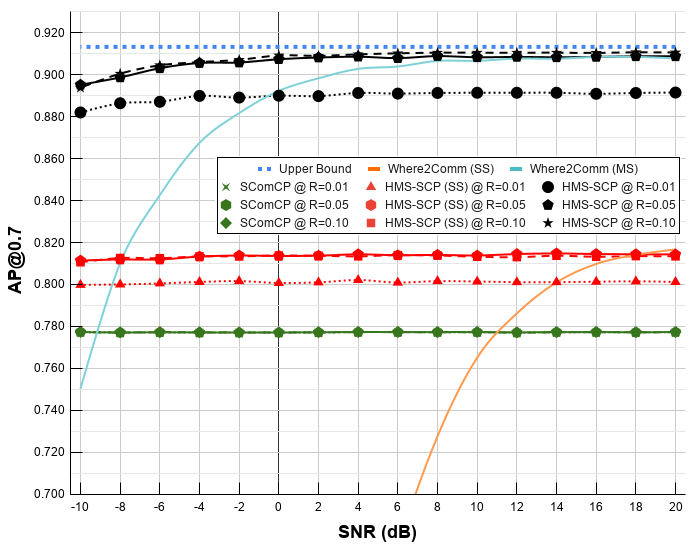}\label{fig:1c}}
    \hfill
    \subfloat[Detection AP@0.7 (Rayleigh)]{\includegraphics[width=0.24\textwidth]{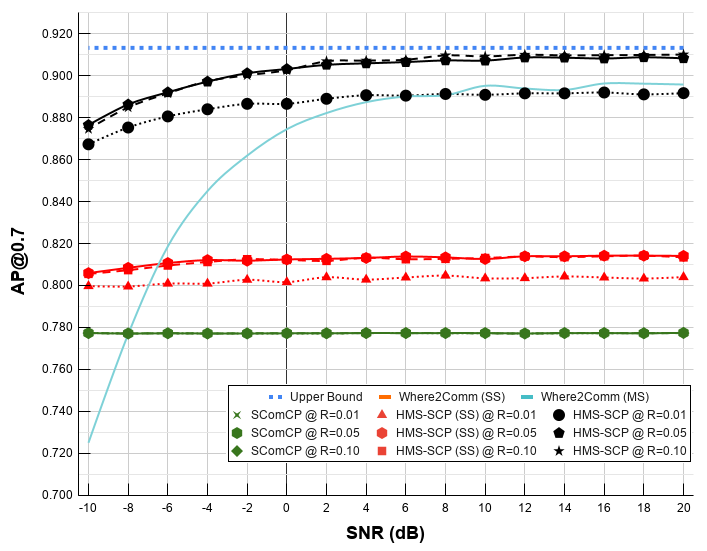}\label{fig:1d}}

    \vspace{-0.5em}

    % Row 2: DAIR-V2X AP@0.5 & AP@0.7
    \subfloat[Detection AP@0.5 (AWGN)]{\includegraphics[width=0.24\textwidth]{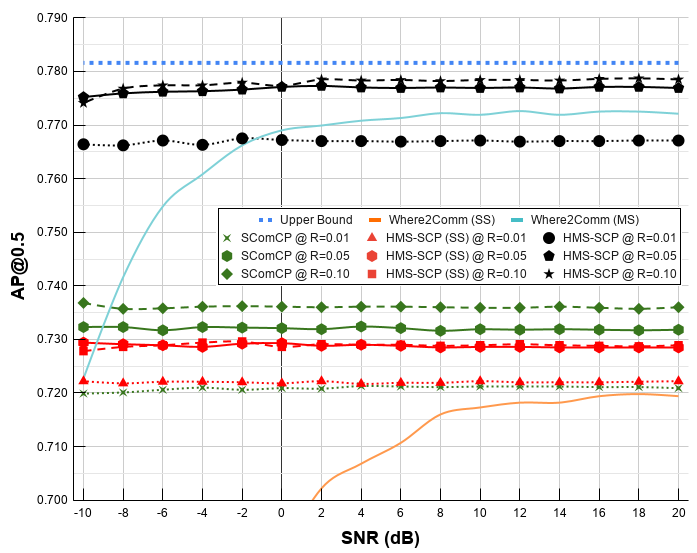}\label{fig:2a}}
    \hfill
    \subfloat[Detection AP@0.5 (Rayleigh)]{\includegraphics[width=0.24\textwidth]{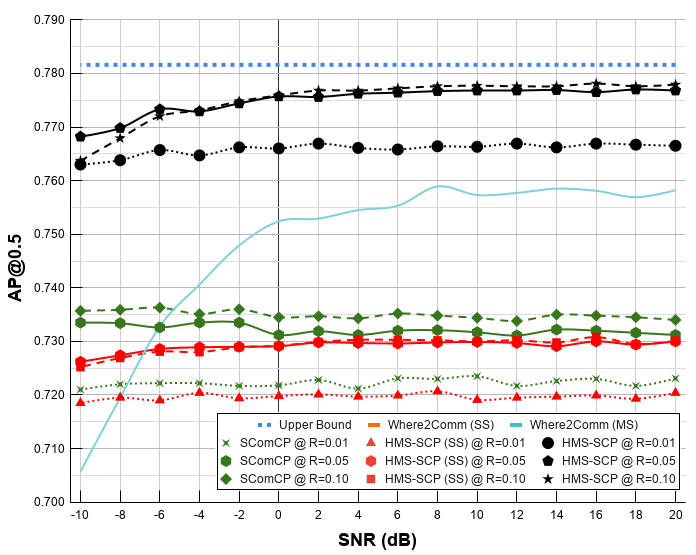}\label{fig:2b}}
    \hfill
    \subfloat[Detection AP@0.7 (AWGN)]{\includegraphics[width=0.24\textwidth]{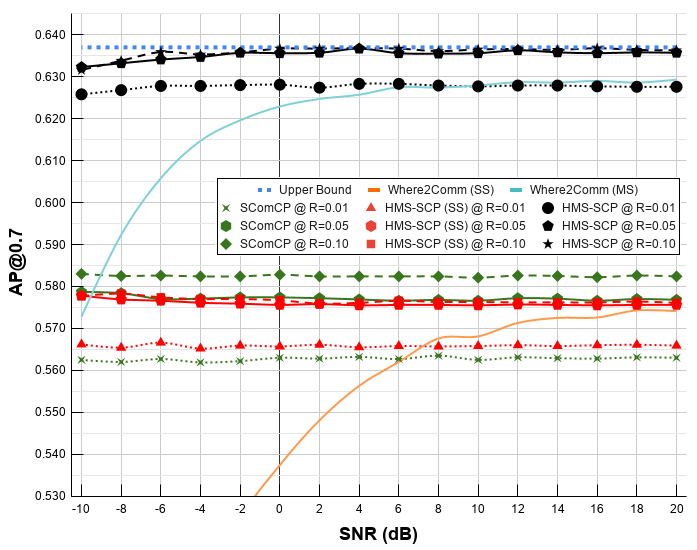}\label{fig:2c}}
    \hfill
    \subfloat[Detection AP@0.7 (Rayleigh)]{\includegraphics[width=0.24\textwidth]{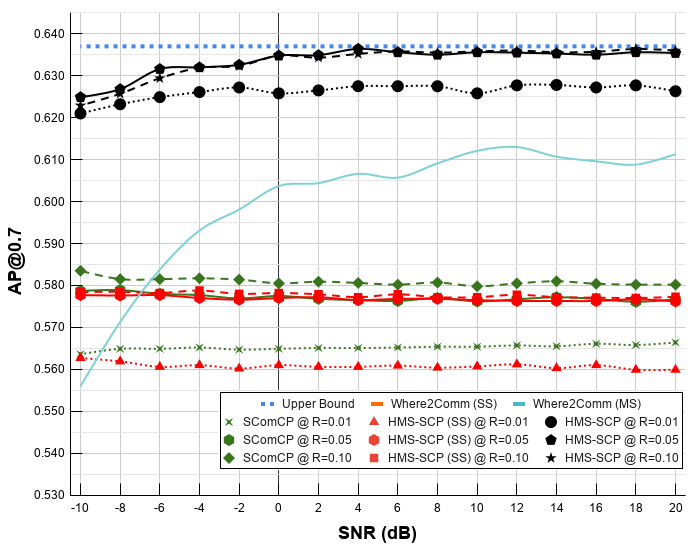}\label{fig:2d}}
    
    \caption{Performance comparison across (a)-(d) OPV2V and (e)-(h) DAIR-V2X datasets under varying SNR conditions.}
    \label{fig:all_perf}
\end{figure*}

\textbf{Quantitative Evaluation:} The performance of the proposed framework under varying SNR conditions on the OPV2V and DAIR-V2X datasets is illustrated in Fig.~\ref{fig:all_perf}. In both figures, the upper bound denotes the maximum achievable performance obtained using the complete semantic feature map under an ideal physical channel. A primary observation is the superior robustness of HMS-SCP under adverse channel conditions compared with non-channel-resilient baselines. Traditional non-channel-resilient single-scale methods like Where2Comm (SS) suffer from a catastrophic cliff effect as SNR drops below 10 dB under the AWGN channel. The degradation becomes more severe under Rayleigh fading, where deep fading and burst-like channel distortions substantially impair the transmitted features. Interestingly, although the uncompressed multi-scale methods like Where2Comm (MS) is capable of achieving high performance under relatively favorable channel conditions, particularly within the 6 to 20~dB range, it still exhibits a pronounced cliff effect at lower SNR levels, typically below 6~dB. This indicates that multi-scale feature representation alone is insufficient to ensure reliable perception unless channel-induced distortions are explicitly considered. 

In contrast, other channel-resilient methods, such as SComCP and our HMS-SCP, maintain a more stable AP performance and exhibit graceful degradation even in challenging Rayleigh fading environments at 0~dB. This noise resilience is primarily attributed to the integration of deep JSCC with task oriented training strategy, which enables semantic features to be optimized jointly for compression, transmission, and perception. Notably, HMS-SCP consistently outperforms both the single-scale baseline SComCP and the single-scale HMS-SCP variant across the two datasets. Despite assigning only one complex-valued symbol per spatial location, resulting in a $256\times$ reduction in overhead compared with SComCP, HMS-SCP achieves a higher perception performance across all ranges for both simulated and real-world datasets. This suggests that hierarchical mapping provides a more effective semantic redundancy than the high-dimensional projection used in previous SOTA methods. Furthermore, the results highlight the significant performance gain provided by the multi-scale representation. At SNR above $10$~dB, HMS-SCP nearly converges to the upper-bound performance, especially using the transmission ratio of $R=0.05$ and $R=0.1$, confirming its ability to achieve a favorable balance between communication efficiency and perception accuracy.
\begin{table*}[t]
\centering
\caption{Reliability Analysis under Challenging Channel Conditions (Rayleigh Fading, $SNR = 0\,\mathrm{dB}$)}
\label{table:reliability_0dB}
\scalebox{0.88}{
\begin{tabular}{@{}lcccccccccccc@{}}
\toprule
& \multicolumn{6}{c}{\textbf{OPV2V Dataset}} & \multicolumn{6}{c}{\textbf{DAIR-V2X Dataset}} \\ \cmidrule(lr){2-7} \cmidrule(lr){8-13}
& \multicolumn{3}{c}{AP@0.5 $\uparrow$} & \multicolumn{3}{c}{AP@0.7 $\uparrow$} & \multicolumn{3}{c}{AP@0.5 $\uparrow$} & \multicolumn{3}{c}{AP@0.7 $\uparrow$} \\ \cmidrule(lr){2-4} \cmidrule(lr){5-7} \cmidrule(lr){8-10} \cmidrule(lr){11-13}
\textbf{Method} & 0--30m & 30--50m & 50--100m & 0--30m & 30--50m & 50--100m & 0--30m & 30--50m & 50--100m & 0--30m & 30--50m & 50--100m \\ \midrule
Where2Comm (SS) & 0.6036 & 0.4695 & 0.3099 & 0.4417 & 0.2718 & 0.1245 & 0.6383 & 0.5532 & 0.3133 & 0.5112 & 0.4255 & 0.1949 \\
Where2Comm (MS) & 0.9856 & 0.9459 & 0.8002 & 0.9660 & 0.8924 & 0.6775 & 0.8502 & 0.7912 & 0.6109 & 0.7449 & 0.6309 & 0.4416 \\
SComCP @ $R=0.01$ & 0.9559 & 0.8989 & 0.6828 & 0.9194 & 0.7819 & 0.4932 & 0.8304 & 0.7690 & 0.5629 & 0.7080 & 0.6160 & 0.3821 \\ \midrule
\textbf{HMS-SCP (SS) @ $R=0.01$} & 0.9795 & 0.9347 & 0.7909 & 0.9483 & 0.8189 & 0.5393 & 0.8342 & 0.7675 & 0.5584 & 0.7000 & 0.6202 & 0.3765 \\
\textbf{HMS-SCP @ $R=0.01$} & \textbf{0.9877} & \textbf{0.9475} & \textbf{0.8170} & \textbf{0.9725} & \textbf{0.9066} & \textbf{0.7041} & \textbf{0.8518} & \textbf{0.8067} & \textbf{0.6344} & \textbf{0.7562} & \textbf{0.6546} & \textbf{0.4726} \\ \bottomrule
\end{tabular}}
\vspace{2pt}
\begin{flushleft}
\scriptsize \textit{Note: Bold values indicate the best performance for the specific range. $R=0.01$ indicates 100$\times$ compression (1 complex-valued symbol per feature).}
\end{flushleft}
\end{table*}

\begin{figure*}[t]
    \centering
    % Left Image: OPV2V
    \begin{minipage}{0.49\textwidth}
        \centering
        \includegraphics[width=\linewidth]{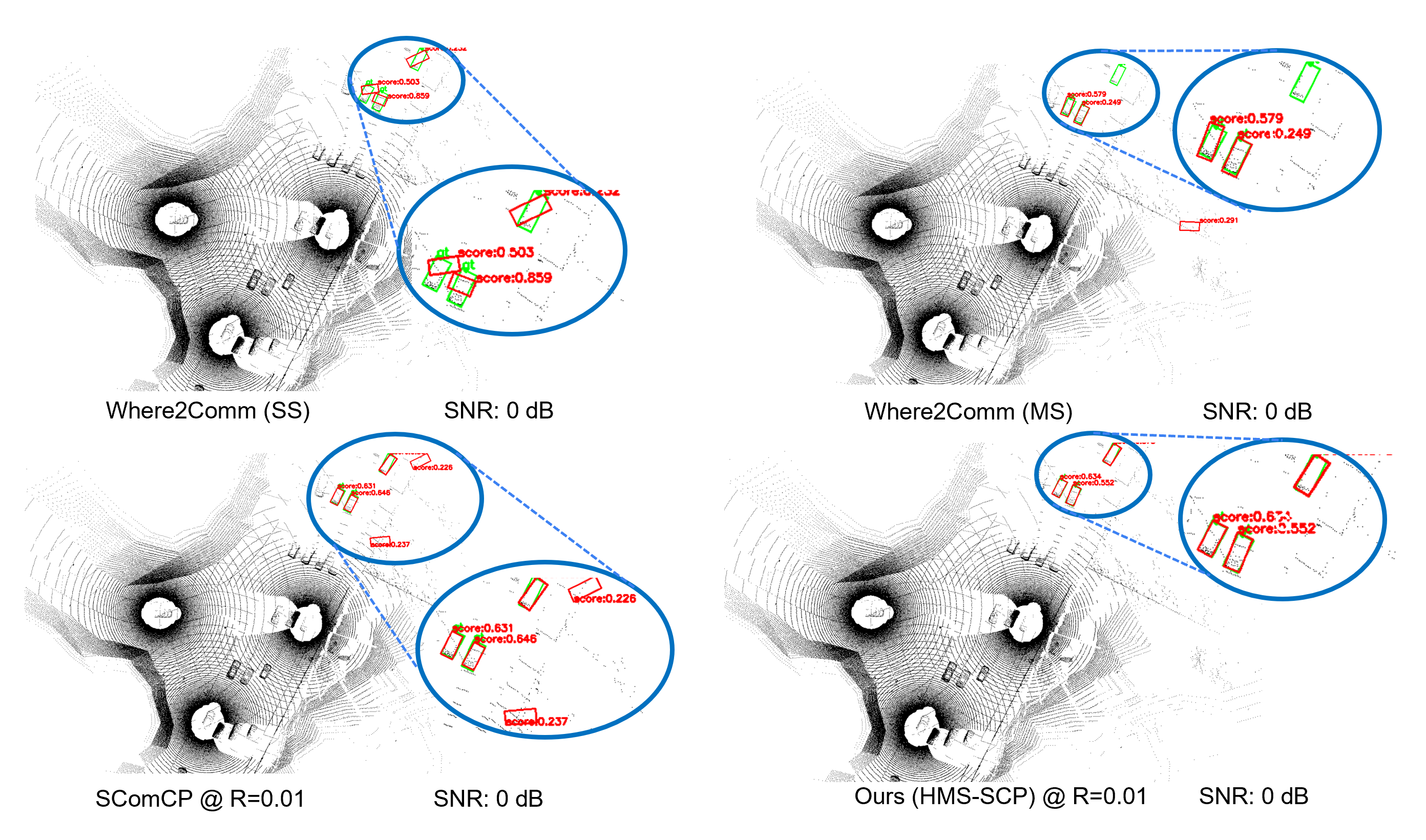}
        \vspace{2pt} % Small gap between image and label
        \centerline{\footnotesize (a) OPV2V Dataset}
    \end{minipage}
    \hfill
    % Left Image: DAIR-V2X
    \begin{minipage}{0.49\textwidth}
        \centering
        \includegraphics[width=\linewidth]{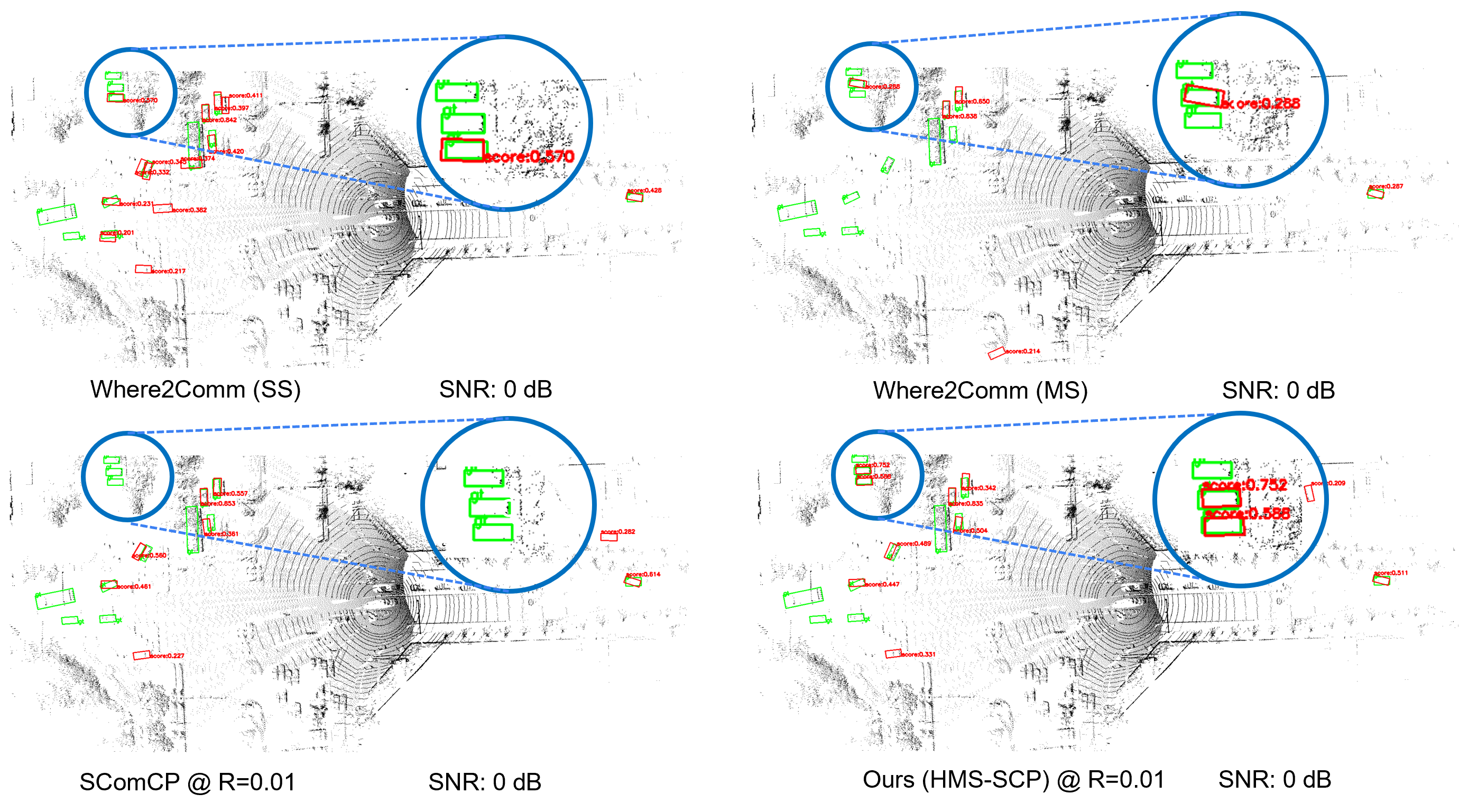}
        \vspace{2pt}
        \centerline{\footnotesize (b) DAIR-V2X Dataset}
    \end{minipage}

    \vspace{4pt} % Gap between sub-labels and main caption
    \caption{Visualization of detection results beyond 50~m under severe Rayleigh fading (SNR = 0~dB) at a 1\% transmission ratio ($R = 0.01$). Green: ground truth; Red: prediction. Our HMS-SCP framework effectively preserves far-field semantic details that are otherwise lost or corrupted in the baseline methods.}
    \label{fig:detection_result}
\end{figure*}

As reported in Table~\ref{table:performance_gap_ms}, HMS-SCP achieves near-upper-bound detection accuracy on the DAIR-V2X dataset even under the challenging 0~dB SNR condition. At the transmission ratio of $R \in \{0.01, 0.05, 0.10\}$, the AP@0.5 performance gaps relative to the upper bound are within $1.99\%$, $0.75\%$, and $0.73\%$, respectively. These results demonstrate that the proposed framework can preserve near-optimal cooperative perception capability while consuming only a small fraction of the available communication budget. Notably, as the transmission ratio increases to $R=0.05$, the performance gap narrows to a negligible $0.75\%$, corresponding to a recovery of $99.25\%$ of the upper-bound detection accuracy. Furthermore, doubling the budget to $R=0.10$ yields a marginal gain of only $0.02\%$, indicating that task-relevant semantic information is largely captured within the first 5\% of the transmitted spatial features. This efficiency is even more pronounced on stricter AP@0.7 metric, where the performance gap remains within $1.76\%$, $0.35\%$, and $0.35\%$ of the upper bound, respectively. The saturation of the AP@0.7 gap at $R=0.05$ further suggests that the core geometric features required for high-precision localization are fully captured within the initial $5\%$ of the transmitted semantic payload. The performance gains are even more prominent under AWGN conditions. A similar comparative analysis for the OPV2V dataset is also provided in Table~\ref{table:performance_gap_ms}. Again, HMS-SCP exhibits remarkable resilience to the challenging $0\text{ dB}$ SNR environment, restricting the AP@0.5 performance gap to within $1.88\%$ of the upper bound even at extreme sparsity ($R=0.01$), demonstrating a near-lossless $0.36\%$ gap under AWGN conditions as the transmission ratio reaches $R=0.10$. These results confirm that HMS-SCP achieves strong robustness across both simulated and real-world datasets, while maintaining high communication efficiency under severe channel constraints. 

As shown in Table~\ref{table:reliability_0dB}, the single-scale baselines experience catastrophic performance collapse in the far-field (50--100m) under severe channel noise, whereas the proposed HMS-SCP architecture maintains a significant perceptual margin. On the DAIR-V2X dataset, the HMS-SCP method outperforms the closest competing method, SComCP, by more than $12.7\%$ in AP@0.5 and a remarkable $23.7\%$ in AP@0.7 at the 100~m horizon, despite utilizing an identical $1\%$ spatial sampling ratio. This performance gain is even more pronounced on the OPV2V dataset, where HMS-SCP exceeds SComCP by $19.7\%$ in AP@0.5 and $42.7\%$ in AP@0.7. These results suggest that hierarchical semantic representations provide a more resilient information bottleneck against channel-induced distortions than traditional high-dimensional projections. In particular, the multi-scale design preserves complementary fine-grained and contextual features that are critical for long-range object detection, thereby improving perception reliability in safety-critical V2X scenarios.

\textbf{Qualitative Evaluation:} Fig.~\ref{fig:detection_result} presents the visualization of detection results for far-field objects beyond 50~m for for both the OPV2V (Fig. \ref{fig:detection_result}a) and DAIR-V2X (Fig. \ref{fig:detection_result}b) datasets. Under severe Rayleigh fading ($SNR = 0~\text{dB}$), all baseline methods exhibit noticeable performance degradation in far-field perception, primarily due to corruption of transmitted semantic features. Specifically, for the DAIR-V2X dataset, SComCP fails to detect any of the three highlighted distant vehicles, whereas HMS-SCP is capable to identify vehicles beyond 50~m with significantly higher confidence score for the true positive predictions. For the OPV2V dataset, HMS-SCP also accurately detects clusters of far-field objects that are either missed or poorly localized by the Where2Comm baselines. Furthermore, HMS-SCP has no false positive predictions compared to SComCP, indicating improved robustness under noisy channel conditions. These qualitative results provide clear evidence that hierarchical semantic representations are more resilient to channel-induced distortions, ensuring perceptual fidelity in safety-critical long-range V2X scenarios.

\subsection{Ablation Studies}

To examine scale-wise contributions of hierarchical feature levels to cooperative perception, a series of ablation experiments are conducted on both datasets under a challenging 0~dB SNR Rayleigh fading channel and an extreme transmission ratio of $R = 0.01$. The far-field results are reported in Table~\ref{table:ablation_studies}. The ablation settings are organized into two categories: transmission suppression, where semantic features from a specific neighboring agent's scale are removed, and fusion ablation, where the model is restricted to ego-only processing at a given scale without using collaborative information from neighboring agents.

The first analysis investigates the importance of each hierarchical scale in the neighbor-to-ego communication channel, with particular emphasis on the 50--100~m horizon, where local sensor limitations are most pronounced. In these configurations, the transmitted feature representation at a specific scale feature $s$ from neighboring agent $j$ is deactivated, forcing the ego agent to rely solely on its local features at that scale. The most significant performance degradation is observed when the neighbor suppresses the transmission of Scale 1. In this case, the AP@0.5 for far-field drops from 0.8170 to 0.6078 for OPV2V and from 0.6344 to 0.5037 for DAIR-V2X. This consistent performance collapse across both datasets confirms that high-resolution semantic data from neighbors is critical for detecting occluded or distant objects that cannot be reliably resolved by the ego agent's own local Scale 1 features. Interestingly, suppressing Scale 3 transmission has a negligible impact on far-field performance in both datasets. This suggests that the coarse global semantic context encoded by Scale 3 is either spatially redundant across agents or can be sufficiently captured by the ego agent's local representation.

Next, multi-scale fusion is evaluated by disabling self-attention at target scales, forcing the ego vehicle to rely strictly on local features at those hierarchical levels. On the DAIR-V2X dataset, the transition from local-only $\mathbf{F}_i^{(1)}$ and $\mathbf{F}_i^{(2)}$ configurations with full HMS-SCP yields substantial absolute gains of 13.6\% and 1.13\% in far-field AP@0.5, respectively. This gain is even more evident on the OPV2V dataset, where disabling Scale 1 fusion causes a notable decline in far-field AP@0.5 to 0.5932. These results indicate that cooperative fusion is particularly critical at high-resolution feature levels, where fine-grained spatial cues are essential for detecting distant objects. Although the local backbone provides a strong baseline representation, the scale-wise fusion mechanism substantially enhances long-range perception by integrating complementary observations from neighboring agents. 

\begin{table}[t]
\centering
\caption{Ablation Analysis: Impact of Hierarchical Scales and Fusion under Rayleigh Fading ($SNR = 0\,\mathrm{dB}$, $R=0.01$)}
\label{table:ablation_studies}
\renewcommand{\arraystretch}{1.2} % Comfortable vertical spacing
\setlength{\tabcolsep}{1.5pt}     % Tight horizontal spacing to fit column
\begin{scriptsize}
\begin{tabular}{@{}lcccccc@{}}
\toprule
 & \multicolumn{3}{c}{\textbf{OPV2V (50--100m)}} & \multicolumn{3}{c}{\textbf{DAIR-V2X (50--100m)}} \\ \cmidrule(lr){2-4} \cmidrule(lr){5-7}
\textbf{Configuration} & \textbf{AP@0.5 $\uparrow$} & \textbf{Precision $\uparrow$} & \textbf{Recall $\uparrow$} & \textbf{AP@0.5 $\uparrow$} & \textbf{Precision $\uparrow$} & \textbf{Recall $\uparrow$} \\ \midrule
\textbf{HMS-SCP (Full)} & \textbf{0.8170} & 0.7683 & \textbf{0.8629} & \textbf{0.6344} & 0.5526 & \textbf{0.6945} \\ \midrule
\textit{Tx Suppression} & & & & & & \\
w/o $\mathbf{z}_j^{(1)}$ & 0.6078 & 0.6856 & 0.6607 & 0.5037 & 0.4905 & 0.5525 \\
w/o $\mathbf{z}_j^{(2)}$ & 0.8035 & \textbf{0.7986} & 0.8463 & 0.6230 & \textbf{0.6031} & 0.6760 \\
w/o $\mathbf{z}_j^{(3)}$ & 0.8167 & 0.7715 & 0.8624 & 0.6346 & 0.5420 & 0.6941 \\ \midrule
\textit{Fusion Ablation} & & & & & & \\
Local-only $\mathbf{F}_i^{(1)}$ & 0.5932 & 0.7921 & 0.6356 & 0.4987 & 0.4789 & 0.5481 \\
Local-only $\mathbf{F}_i^{(2)}$ & 0.8185 & 0.6852 & 0.8697 & 0.6231 & 0.5406 & 0.6800 \\
Local-only $\mathbf{F}_i^{(3)}$ & 0.8161 & 0.7688 & 0.8621 & 0.6347 & 0.5435 & 0.6941 \\ \midrule
\textbf{HMS-SCP (SS)} & 0.7942 & 0.6865 & 0.8629 & 0.5590 & 0.5024 & 0.6421 \\ \bottomrule
\end{tabular}
\end{scriptsize}
\end{table}

Lastly, far-field precision and recall metrics are analyzed to evaluate performance from a safety-critical perspective. On both datasets, suppressing Scale 2 transmission produces the highest far-field Prec@0.5, reaching 0.7986 for OPV2V and 0.6031 for DAIR-V2X. However, this gain in precision is achieved at the expense of Rec@0.5, which drops to 0.8463 and 0.6760, respectively. In C-ITS context, false negatives, such as missed vehicles, are generally more safety-critical than marginal gains in bounding-box precision, as they directly affect hazard awareness and reaction time. Thus, the full hierarchical design remains preferred, achieving the highest far-field Rec@0.5 (0.8629 on OPV2V, 0.6945 on DAIR-V2X). This demonstrates superior target coverage in the safety-critical 50--100~m range, offering an optimal precision-recall balance for long-range object awareness.

Overall, the consistency in hierarchical dependencies observed on both the simulated OPV2V and real-world DAIR-V2X datasets confirms that the proposed HMS-SCP is broadly effective across diverse cooperative perception settings. When trained and evaluated under the standard per-dataset protocol, the framework delivers stable performance gains on both datasets, indicating strong architectural adaptability and consistent effectiveness. These results show that the hierarchical multi-scale approach can balance semantic richness and communication overhead across different data distributions and sensor configurations, highlighting its versatility for various V2X scenarios.

\subsection{Bandwidth and Communication Efficiency Analysis}

In this section, the communication efficiency of the proposed HMS-SCP is quantified to evaluate its feasibility for real-world V2X deployment. HMS-SCP applies a fixed transmission ratio $R=0.01$, corresponding to a 1\% spatial sampling rate, to the spatial grid points of each hierarchical scale. Each selected spatial location is then mapped to exactly one complex-valued symbol, effectively compressing the corresponding multi-channel semantic latent into a single transmission element. As detailed in Table \ref{table:bandwidth_analysis}, this design results in a lightweight footprint of only 331 symbols for the DAIR-V2X dataset and 444 symbols for OPV2V dataset per perception cycle. The hierarchical structure normally maintains a 16:4:1 symbol ratio across scales, with the majority of the bandwidth dedicated to high-resolution Scale 1 features, which constitutes 76.2\% of the total bandwidth budget. This allocation is consistent with the ablation findings, where Scale 1 features are shown to be most critical for long-range perception. In contrast, Scale 3 requires only 16 to 21 symbols, reflecting its role in providing compact global structural context with minimal communication overhead.

\begin{table}[h]
\centering
\caption{Lightweight Hierarchical Symbol Footprint of HMS-SCP ($R=0.01$)}
\label{table:bandwidth_analysis}
\renewcommand{\arraystretch}{1.2}
\resizebox{\columnwidth}{!}{%
\begin{tabular}{@{}lcccc@{}}
\toprule
\multirow{2}{*}{\textbf{Scale}} & \multicolumn{2}{c}{\textbf{OPV2V}} & \multicolumn{2}{c}{\textbf{DAIR-V2X}} \\ \cmidrule(lr){2-3} \cmidrule(lr){4-5} 
 & \textbf{Grid ($H \times W$)} & \textbf{Symbols} & \textbf{Grid ($H \times W$)} & \textbf{Symbols} \\ \midrule
Scale 1 & $96 \times 352$ & 338 & $100 \times 252$ & 252 \\
Scale 2 & $48 \times 176$ & 85 & $50 \times 126$ & 63 \\
Scale 3 & $24 \times 88$ & 21 & $25 \times 63$ & 16 \\ \midrule
\textbf{Total} & --- & \textbf{444} & --- & \textbf{331} \\ \bottomrule
\end{tabular}%
}
\end{table}

The proposed HMS-SCP offers a more communication-efficient architecture than the SComCP baseline\cite{11355867}. SComCP transmits an average of 12 selected semantic features protected by a high symbol expansion with 256 symbols per feature, resulting in approximately 3,072 symbols per perception cycle in each communication round. While SComCP consumes nearly 7 times more bandwidth than the proposed HMS-SCP, it substantially limits the number of spatial locations that can be communicated, leading to a highly sparse spatial representation. In contrast, HMS-SCP adopts a high-efficiency 1-to-1 semantic mapping strategy by relocating approximately 37 times (444 vs. 12) or 27 times (331 vs. 12) more spatial anchors in the environment. This spatially-augmented approach extending the grid density allowing the model to better capture distant vehicles that sparse models may overlook. As a result, this provides an optimal balance between transmission volume and perceptual reliability. This remarkably small footprint of 331 or 444 symbols per perception cycle underlines the practical feasibility of HMS-SCP for real-world V2X deployment. Considering that 3GPP C-V2X operates in the ITS 5.9~GHz spectrum, the required transmission payload constitutes significantly less than 1\% of the available capacity per perception cycle. Under a standard 10~MHz channel and a typical LiDAR operating at 10~Hz, the proposed symbol footprint constitutes merely 0.04\% of the available channel capacity per cycle. This efficiency ensures that HMS-SCP can scale to dense traffic without overloading the channel, while meeting the low-latency demands of safety-critical cooperative perception.

\subsection{Complexity and Latency Analysis}

To evaluate the practical feasibility of the proposed framework for real-time V2X applications, a comprehensive complexity and latency analysis is conducted. The computational overhead is quantified in terms of learnable parameters, while inference latency is measured on the OPV2V and DAIR-V2X benchmarks. The results are summarized in Table~\ref{table_complexity}. The proposed HMS-SCP architecture comprises 8.70M learnable parameters. As shown in Table~\ref{table_complexity}, the overall complexity of the architecture is primarily dominated by the feature extraction module, which accounts for $68.72\%$, corresponding to 5.98M parameters. Notably, the additional modules introduced by the HMS-SCP framework remain highly lightweight; the feature selection mechanism implemented through the spatial importance predictor requires only 129K parameters ($1.49\%$), and the semantic codec contributes an additional 514K parameters ($5.90\%$). This limited parameter overhead shows that the proposed modules introduced a minimal memory footprint, making the framework suitable for deployment on resource-constrained V2X edge devices.

In addition to detection accuracy, real-time computational efficiency is essential for safety-critical cooperative perception\cite{10274112}. The proposed HMS-SCP framework achieves an end-to-end inference latency of 15.73~ms on the OPV2V dataset and 12.17~ms on the DAIR-V2X dataset, which translates to approximately 63 and 82 frames per second (FPS), respectively. Detailed profiling of the pipeline runtime reveals that the feature fusion module remains the primary computational bottleneck, requiring 4.95--6.73~ms due to intensive spatial alignment and coordinate transformation operations performed across multiple agents. Critically, the newly introduced HMS-SCP components, such as feature selection and semantic codec, introduce only marginal overhead, collectively contributing approximately 4--5~ms to the overall pipeline. Given that total processing time remains significantly below the 100~ms latency threshold mandated for safety-critical V2X applications\cite{3gpp_ts_22186, 10.1145/3296957.3173191}, these results confirm that HMS-SCP provides sufficient throughput for high-speed collaborative 3D detection in dynamic and complex urban environments.

\begin{table}[!t]
\centering
\caption{Complexity Analysis with Latency}
\label{table_complexity}
\addtolength{\tabcolsep}{-3pt}
\begin{tabular}{lcccc}
\toprule
\textbf{Components} & \textbf{Params} & \textbf{\%} & \textbf{OPV2V} & \textbf{DAIR-V2X} \\
 & & & \textbf{(ms)} & \textbf{(ms)} \\
\midrule
Feature Extraction       & 5.98M & 68.72 & 3.43  & 3.30 \\
Feature Selection     & 129K  & 1.49  & 1.26  & 0.90 \\
Semantic Codec        & 514K  & 5.90  & 4.13  & 2.88 \\
Feature Fusion     & 2.07M & 23.83 & 6.73  & 4.95 \\
Detection Head     & 5.1K  & 0.06  & 0.18  & 0.14 \\
\midrule
\textbf{Total}     & \textbf{8.70M} & \textbf{100} & \textbf{15.73} & \textbf{12.17} \\
\bottomrule
\end{tabular}
\end{table}

\section{Conclusion}

In this paper, we propose HMS-SCP, a hierarchical multi-scale semantic-aware cooperative perception framework for task-oriented semantic communication in V2X collaborative 3D object detection. Unlike existing frameworks that typically operate on a single intermediate feature map, HMS-SCP exploits a spatial importance predictor that dynamically identifies critical semantic features across hierarchical feature levels, which are then directly mapped to complex-valued symbols via a JSCC codec to streamline the communication bottleneck. This multi-scale selection mechanism enables the system to dynamically allocate bandwidth across different semantic levels, prioritizing fine-grained details or global context depending on their contribution to the perception task. To address the high uncertainty of V2X communication in dense traffic, HMS-SCP leverages structural multi-scale redundancy across multiple feature scales rather than relying on high-dimensional symbol projections. This allows the framework to achieve an ultra-low symbol rate while mitigating network congestion and preserving perception reliability. 

Extensive experiments on the simulated OPV2V dataset and the real-world DAIR-V2X dataset demonstrate that HMS-SCP consistently outperforms SOTA baselines under both AWGN and Rayleigh fading channels. Under severe Rayleigh fading, HMS-SCP outperforms the closest competing method by $23.7\%$ on DAIR-V2X and $42.7\%$ on OPV2V at the far-field range beyond 50~m horizon. This has confirmed the effectiveness of hierarchical semantic transmission for long-range cooperative perception. While the current framework employs a uniform spatial compression ratio across all hierarchical scales, future research will explore adaptive, scale-wise ratio allocation to further optimize the trade-off between communication efficiency and perceptual robustness. Furthermore, extending HMS-SCP to multi-modal cooperative perception by incorporating camera, radar, and LiDAR sensors represents a promising direction for achieving more accurate and resilient perception in highly dynamic vehicular environments.

\bibliographystyle{IEEEtran}
\bibliography{reference}
\newpage

\vfill
\end{document}